\documentclass[10pt,oneside,onecolumn]{elsarticle}

\usepackage[ansinew]{inputenc}
\usepackage[english]{babel}
\usepackage{ifthen,shortvrb}
\usepackage[fleqn]{amsmath}
\usepackage{amsxtra,amsfonts,latexsym,amssymb,euscript}
\usepackage{amstext}
\usepackage{graphicx}
\usepackage{dcolumn}
\usepackage{subfigure}
\usepackage{mathrsfs}
\usepackage{arydshln}
\usepackage{algorithm}

\newcommand{\bbu}{\bar{\mathbf{u}}}
\newcommand{\bu}{\mathbf{u}}
\newcommand{\bp}{\mathbf{p}}
\newcommand{\bH}{\mathbf{H}}
\newcommand{\IC}{\mathcal{I}}

\newcommand{\II}{\mathcal{I}_0}

\begin{document}
\begin{frontmatter}
\journal{Wear}
\title{Evolution of the free volume between rough surfaces in contact}
\author[a,b]{M. Paggi\corref{cor1}}
\ead{marco.paggi@imtlucca.it}
\author[a]{Q.-C. He}
\ead{qi-chang.he@univ-paris-est.fr}
\cortext[cor1]{Corresponding author. Tel: +39-0583-4326-604, Fax: +39-0583-4326-565}
\address[a]{Universit\'{e} Paris Est, Laboratoire Mod\'{e}lisation
et Simulation Multi Echelle, MSME UMR 8208 CNRS, 5 Bd Descartes,
77454 Marne-la-Vall\'{e}e Cedex 2, France}
\address[b]{IMT Institute for Advanced Studies Lucca, Piazza San Francesco 19, 55100 Lucca, Italy}

\begin{abstract}
The free volume comprised between rough surfaces in contact governs
the fluid/gas transport properties across networks of cracks and the
leakage/percolation phenomena in seals. In this study, a fundamental
insight into the evolution of the free volume depending on the mean
plane separation, on the real contact area and on the applied
pressure is gained in reference to fractal surfaces whose contact
response is solved using the boundary element method. Particular
attention is paid to the effect of the surface fractal dimension and
of the surface resolution on the predicted results. The free volume
domains corresponding to different threshold levels are found to
display fractal spatial distributions whose bounds to their fractal
dimensions are theoretically derived. A synthetic formula based on
the probability distribution function of the free volumes is
proposed to synthetically interpret the numerically observed trends.

\vspace{1em} \noindent Notice: this is the authors version of a work
that was accepted for publication in Wear. Changes resulting from
the publishing process, such as editing, structural formatting, and
other quality control mechanisms may not be reflected in this
document. A definitive version was published in Wear, Vol. 336,
86--95, DOI:10.1016/j.wear.2015.04.021
\end{abstract}

\end{frontmatter}

\section{Introduction}

Contact mechanics between rough surfaces is a topic of paramount
importance in engineering and physics, since surface phenomena in
nature and technology strongly depend on the topological properties
of interfaces. Real surfaces are never ideally flat and roughness is
present at different scales, from the specimen size down to the
interatomic distance. Hence, when two bodies are pressed against
each other, contact takes place at the asperities (the 3D maxima of
the surfaces) and the real contact area is a fraction of the nominal
one.

In the context of rough surfaces, the scientific community has paid
particular attention to the relation between the real contact area
and the applied pressure \cite{GW66,MB,BCC,persson,CB08,PC10,YAM},
the contact stiffness
\cite{CMY,Ciavarella1,Ciavarella2,Ciavarella3,RP1,RP2} which is
proportional to the electric and thermal contact conductances
\cite{barber,PB,r1,r2}, frictional phenomena \cite{BCC01,PPP},
adhesion \cite{carbone1,carbone2}, and hydrophobic properties of
surfaces \cite{carbone3,noson}. Recent studies on rough surfaces
have also elasto-plastic contact \cite{r3,r4}, adhesive contact
\cite{r5}, and lubrication \cite{r6}.

Another important topic regards the transport properties of rough
surfaces in contact. Below the full contact limit, a free volume
between the contacting bodies is always present due to roughness.
Such a free volume constitutes a fractal network whose properties
are important for flow and transport of hydrothermal fluids, water,
and contaminants in groundwater systems, but also of oil and gas in
petroleum reservoirs \cite{BB02}. For instance, the transport
properties of proppant through fracture networks are relevant for
hydraulic fracturing \cite{antoun}. At a much smaller scale, welded
surfaces in micro-electro-mechanical systems (MEMS) may present a
free volume forming channels and capillaries of random distribution.
Such channels are critical for gas leakage that may penetrate the
soldered joint and affect the reliability of the system \cite{han}.
These problems are also relevant in materials for energy
applications, such as in solid oxide fuel cells \cite{green} and in
photovoltaic modules where humidity can diffuse along the interface
between the textured surface of solar cells and the encapsulating
polymer, promoting a chemical degradation of electric contacts. The
topological features of roughness in seal contacts are also very
important for the onset of wear, see \cite{wear1}.

Attempts to predict the transport properties across these finite
thickness interface regions composed of voids and contact areas are
relatively recent and rely on the theory of fractal porous media
\cite{YC,ZXYY}. Based on this modelling assumption, simplified
theories are put forward where the free volumes are treated as pores
of spherical shape with diameter obeying a power-law distribution.
Pioneering analytical models have been proposed in
\cite{epl1,epl2,bottiglione} by examining the evolution of the
contact area depending on the surface resolution. For a flat
surface, the full contact regime takes place and no percolation
channels are present. By refining the surface resolution, roughness
comes into play and the real contact area becomes a fraction of the
nominal one. For a given critical resolution, a first percolating
channel will be originated. Further surface refinements will lead to
other percolation channels that may contribute to the global leakage
rate. Such contributions have been neglected in
\cite{epl1,epl2,bottiglione}. Due to such simplifying assumption,
predictions were found in good agreement with experimental results
only in the low pressure regime.

A rigorous computational approach to predict the contact area and
the transmissivity and diffusivity of the network of the created free
channels was recently proposed in \cite{VLSZ}, where the problem was tackled from the numerical point of view by
using the boundary element method (BEM). However, the analysis was restricted to two specific
surface topologies created by lapping or sand blasting treatments and general trends were not discussed.

In the present study we propose an extensive numerical investigation
of the evolution of the free volume between fractal rough surfaces
in contact with an elastic half plane as a function of the main
contact variables, i.e., the mean plane separation, the contact
force and the real contact area. A computational approach based on
BEM, analogous to that described in \cite{VLSZ}, is used. A deep
analysis of the morphological properties of the free volume domains
is performed, without making simplifying assumptions \emph{a priori}
on their shape and distribution, as in previous models based on the
percolation theory. Moreover, all the channels are considered
without any approximation apart from that arising from the spatial
discretization intrinsic in the method. The obtained numerical
trends and their interpretation are expected to provide useful hints
for the development of further semi-analytical models taking into
account the observed scaling laws, or to refine the existing ones.

The article is structured as follows. In Section 2, the numerical
method used to generate the rough surfaces is outlined and the
fundamental equations of the boundary element method used to solve
the contact problem are described. In Section 3, numerical results
are presented and focus on the scaling of the free volume and on the
multi-scale characterization of its network pattern. Further
theoretical considerations on the statistical distribution of the
free volumes are provided in Section 4, along with a synthetic
formula for the computation of the free volume and for a deeper
understanding of the observed numerical trends. Conclusions and
outlook on the relevance of the proposed methodology for the study
of wear in seal applications complete the study.

\section{Numerical method}

Rough surfaces with fractal properties are numerically generated
according to the random midpoint displacement (RMD) algorithm
\cite{PS}. This method allows generating rough surfaces with a power
spectral density function of power-law type, characterized by a
given fractal dimension $D$ $(2<D<3)$. Applications of the method to
contact mechanics can be found in \cite{BCC,PC10,ZBP}. Square
surfaces with different resolutions can be generated by successively
refining an initial mesh by a successive addition of a series of
intermediate heights. In the algorithm, the number of successive
refinements is defined by the parameter $m$, which is related to the
number of heights per side of the squared generated grid, $2^m+1$.
Given $L$ the lateral size of the surface, the grid spacing is
$\delta=L/2^m$ and the resolution can be defined as $s=1/\delta$.
The method generates surfaces with higher $m$ that are finer
representations of the coarser ones, i.e., the height field of a
surface with $m=i$, $i\in \mathbb{N}$, contains the height field of
the coarser realizations with $m<i$.

\begin{figure}
\centering
\includegraphics[width=.9\textwidth,angle=0]{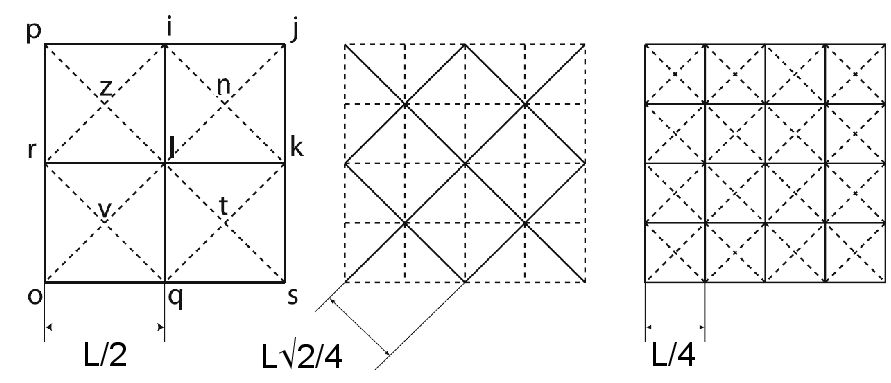}
\caption{Recursive steps for the generation of rough surfaces using
the RMD algorithm.}\label{fig1}
\end{figure}

A sketch showing how the RMD algorithm operates is provided in
Fig.\ref{fig1}. Starting with $m=1$, the elevation of the four
corner nodes of the grid, nodes $o$, $p$, $j$, $s$ in
Fig.\ref{fig1}, are set equal to zero. Afterwards, the elevation of
the central point of the grid, $l$, is determined by the average
value of the elevations of the corner nodes, plus a random number
extracted from a Gaussian distribution with zero mean and variance
$\sigma_1^2=\sigma_0^2/2^{(3-D)/2}$, where $\sigma_0^2$ is a free
parameter set equal to $1/\sqrt{0.09}$. The elevations of the nodes
$i$, $k$, $q$, $r$ are then assigned by averaging over three
elevations, those of the two corner nodes and that of the central
node, plus a random number extracted from a Gaussian distribution
with zero mean and reduced variance
$\sigma_2^2=\sigma_1^2/2^{(3-D)/2}$. This procedure is further
iterated at the next refinement, $m=2$. This version differs from
the original RMD algorithm detailed in \cite{PS} by the fact that
the elevations of the four initial corner nodes are set equal to
zero rather than randomly assigned. The reason is to avoid to create
topologies dominated by these initial values, which might constitute
a bias especially at low resolution.

The most accurate solution of the contact problem between the
generated rough surfaces and a smooth plane by keeping as minimum as
possible the simplifying assumptions on the surface geometry can be
achieved by using the boundary element method for contact mechanics
\cite{KJ}. By imposing a far-field closing displacement $\Delta$ to
the bodies in contact, the displacement at each point of the contact
area is related to the contact pressures as follows \cite{jrB}:
\begin{equation}
u(\mathbf{x})=\int_S H(\mathbf{x},\mathbf{y})p(\mathbf{y})\mbox{d}S,
\end{equation}
where $u(\mathbf{x})$ is the displacement at the surface point
defined by the position vector $\mathbf{x}$,
$H(\mathbf{x},\mathbf{y})$ is the displacement at $\mathbf{x}$ due
to a unit pressure acting at $\mathbf{y}$, and $S$ is the apparent
contact area. Assuming linear elastic isotropic materials, the
influence coefficients are given by \cite{KJ,jrB}:
\begin{equation}
H(\mathbf{x},\mathbf{y})=\dfrac{1-\nu^2}{\pi
E}\dfrac{1}{\parallel\mathbf{x}-\mathbf{y}\parallel},
\end{equation}
where $E$ and $\nu$ denote, respectively, the composite Young's
modulus and Poisson's ratio of the materials of the bodies in
contact. Upon discretization of the surface as a grid where each
nodal height defined by the indices $(i,j)$ is modelled as a square
punch with an elevation $\xi_{i,j}$ above a reference plane
coincident with the level of the smallest surface height, the
contact problem requires the simultaneous solution of the following
set of equations and the satisfaction of a set of inequalities:
\begin{subequations}
\begin{align}
&u_{i,j}=\sum_{k=1}^{N_c}\sum_{l=1}^{N_c}H_{i-k,j-l}\,p_{k,l},\quad
1\le i\le N_c,\; 1\le j\le N_c\\
&u_{i,j}=\bar{u}_{i,j}=\Delta-\xi_{i,j},\quad\forall (i,j)\in \IC\\
&p_{i,j}>0,\quad\forall (i,j)\in \IC\\
&u_{i,j}>\bar{u}_{i,j},\quad\forall (i,j)\in \bar\IC\\
&p_{i,j}=0,\quad\forall (i,j)\in \bar\IC
\end{align}
\end{subequations}
where $\IC$ is the domain of boundary elements in contact, $\bar\IC$
is the domain of boundary elements not in contact, and $\Delta$ is
the imposed far-field displacement. An initial (trial) contact
domain is chosen as the set $\II$ of boundary elements that
compenetrate in the half-plane in case of a rigid body motion, i.e.,
by neglecting the deformation induced by elastic interactions. The
number of boundary elements belonging to this set is $N_c$. The
optimal contact domain is found iteratively by a suitable
elimination of the points bearing tensile (negative) forces
\cite{KJ,BP}. During contact, the evolution of the gaps between the
surface grid points and the half-plane as a result of the elastic
interactions is traced. From the computed normal displacements, the
free volume of each boundary element is
$v_{i,j}=(u_{i,j}-\bar{u}_{i,j})\delta^2$. All the points belonging
to the contact domain, $\IC$, have by definition
$u_{i,j}=\bar{u}_{i,j}$ and therefore $v_{i,j}=0$. The total free
volume $V$ is finally evaluated by summing up the individual
boundary element contributions.

The main steps of the contact algorithm are summarized in Algorithm
1, where $\bH$ denotes the matrix collecting the influence
coefficients, the vector $\bbu$ collects the imposed displacements,
and the vector $\bp$ collects the contact forces. The boundary
elements subject to tensile forces are eliminated from the contact
domain at the step (2.2.1). After convergence, the optimal solution
is stored in the vector $\bp_{\rm{opt}}$ in the step (3). The set of
boundary elements not in contact, $\bar\IC$, is updated in the step
(4). The displacements of the whole surface are computed in the step
(5) and they are equal to $\bbu_\IC$ for the elements in contact,
whereas they are given by the elastic equations for the elements not
in contact, using the corresponding flexibility matrix
$\bH_{\bar\IC,\IC}$.

The output is given by the following dimensionless quantities: the
real contact area fraction, $A^*:=A/A_n$, where $A_n=L^2$; the
dimensionless pressure, $p^*:=p L/(E\sigma)$, where $E$ is the
composite Young’s modulus of the contacting bodies and $\sigma$ is
the r.m.s. of the distribution of surface heights, also called $S_q$
in \cite{standard1,standard2}; the dimensionless free volume,
$V^*:=V/(L^2\sigma)$.

The above dimensionless formulation has been chosen in order to have
results invariant with respect to a transformation of the in plane
coordinates of the type $x\rightarrow\lambda x$,
$y\rightarrow\lambda y$ $(\lambda \in \mathbb{R})$. Such a linear
mapping leads to $L\rightarrow\lambda L$, $p\rightarrow p/\lambda$,
and $V\rightarrow \lambda^2 V$. Hence, the dimensionless pressure
$p^*$ and the dimensionless volume $V^*$ defined as before are not
affected by this type of scaling. However, it has to be pointed out
that this is true if and only if asperities deform linear
elastically. Hence, a lower bound to the surface dimension $L$ that
can be explored for a given number of heights per side without the
occurrence of plastic deformation does exist. Such a critical
lateral size, $L_c$, can be identified for a given material by
setting the plasticity index equal to unity. Using for instance the
expression proposed by Greenwood and Williamson \cite{GW66},
$\psi=\sqrt{\sigma/\rho}E/H$, where $H$ is the material hardness and
$\rho$ is the average radius of curvature of the asperities (which
is the only parameter entering $\psi$ affected by the linear
mapping), we obtain the critical radius of curvature
$\rho_c=\sigma(E/H)^2$ below which asperities are expected to deform
plastically. Therefore, any scaling of the type $L\rightarrow
\lambda L$ with $\lambda<1$ and leading to $\rho\rightarrow
\lambda\rho$ leads to the same $p^*$ and $V^*$ computed for
$\lambda=1$ provided that $\rho>\rho_c$. For $\rho\le\rho_c$, the
present computational model should be extended by taking into
account plastic deformations.

\begin{algorithm}
\caption{Contact algorithm for a given imposed far-field
displacement $\Delta$} \label{1} \vspace*{.1cm}\hrule\vspace*{.1cm}
~~\textbf{Input}: Matrix $\bH$, vector $\bbu$, initial guess $\bp$,
initial active set $\II$, maximum number $K_{\rm max}$ of
iterations, tolerance $\epsilon$. \vspace*{.1cm}\hrule\vspace*{.1cm}
\begin{enumerate}
\item $k\leftarrow 0$, $\IC\leftarrow \II$;
\item \textbf{while} ($k\leq K_{\rm max}$ \textbf{and} $\min(\bp)<-\epsilon$)
\textbf{or} $k=0$ \textbf{do}:
\begin{enumerate}
\item [(2.1)] Solve the unconstrained system of equations
$\bH\bp=\bbu$ for $\bp$ using the Gauss-Seidel algorithm
\item [(2.2)]\textbf{for} $(i,j)\in\IC$ \textbf{do}:
\begin{enumerate}
\item [(2.2.1)]\textbf{if} $p_{i,j}<-\epsilon$ \textbf{then} $p_{i,j}\leftarrow 0$;
$\IC\leftarrow\IC\setminus\{(i,j)\}$;
\end{enumerate}
\end{enumerate}
\item $\bp_{\rm{opt}}\leftarrow \bp$;
\item $\bar\IC\leftarrow\{(1,\ldots,i,\ldots,N_c)\times(1,\ldots,j,\ldots,N_c)\}\setminus\IC$;
\item $\bu_\IC=\bbu_\IC$, $\bu_{\bar\IC}\leftarrow
\bH_{\bar\IC,\IC}\bp_\IC$;
\item \textbf{end}.
\end{enumerate}
\vspace*{.1cm}\hrule\vspace*{.1cm} ~~\textbf{Output}: dimensionless
nominal pressure $p^*$, dimensionless contact area $A^*$, and
dimensionless free volume $V^*$. \vspace*{.1cm}
\end{algorithm}

\section{Numerical results}

\subsection{Scaling of the real contact area and of the free volume}

In the contact between bodies with fractal boundaries and
multi-scale roughness, previous research \cite{MB,BCC,persson,ZBP}
has highlighted the important role of the fractal dimension and of
the surface resolution (lower cut-off length) on the relation
between the real contact area and the contact pressure. Moreover,
the contact domain was topologically characterized in \cite{BCC} and
it was found to be of lacunar type, with fractal properties
dependent on those of the parent undeformed fractal surface and on
the applied pressure.

To investigate how the free volume between rough surfaces in contact
scales with the real contact area and with the contact pressure, we
herein perform a series of contact simulations by using the
numerical method described in the previous section. All the
numerical tests are performed under displacement control. A
displacement $\Delta$ is imposed in the normal direction to the flat
contacting plane. The maximum value of $\Delta$ is equal to the
difference between the elevation of the highest asperity and the
elevation of the mean plane of the asperity heights computed in the
undeformed configuration. This total displacement is subdivided in
60 steps.

For each imposed normal displacement, the grid points in contact are
determined and the real contact area $A$ and the contact pressure
$p$ computed by summing up the contributions of the individual
square punches. The volume $V$ comprised between the flat plane and
the rough surface is finally determined from the computed
displacement field and the original surface geometry according to
the method described in Section 2. Numerical results are interpreted
via the following dimensionless quantities: the real contact area
fraction, $A^*=A/A_n$; the dimensionless pressure, $p^*=p
L/(E\sigma)$; the dimensionless free volume, $V^*=V/(L^2\sigma)$.

\begin{figure}
\centering
\subfigure[$D=2.1$]{\includegraphics[width=.32\textwidth,angle=0]{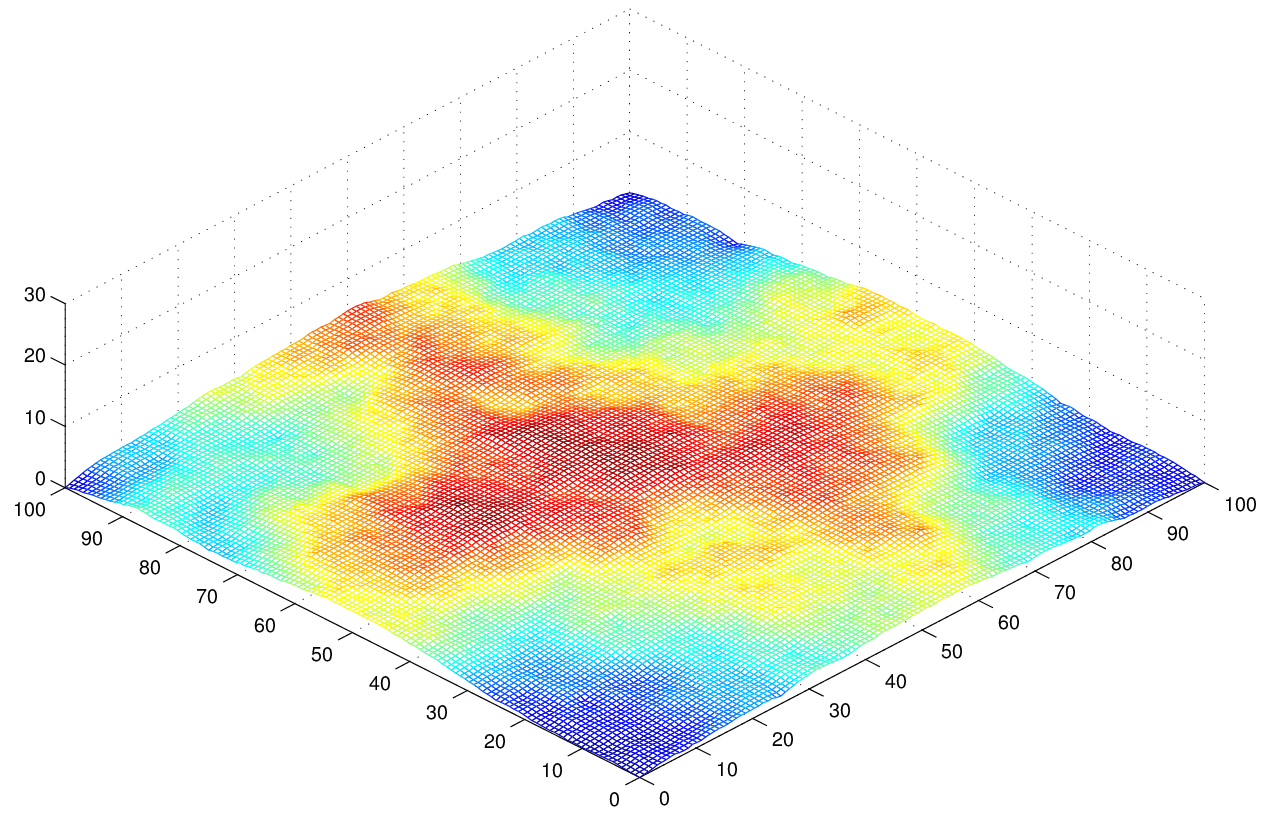}\label{fig2a}}
\subfigure[$D=2.5$]{\includegraphics[width=.32\textwidth,angle=0]{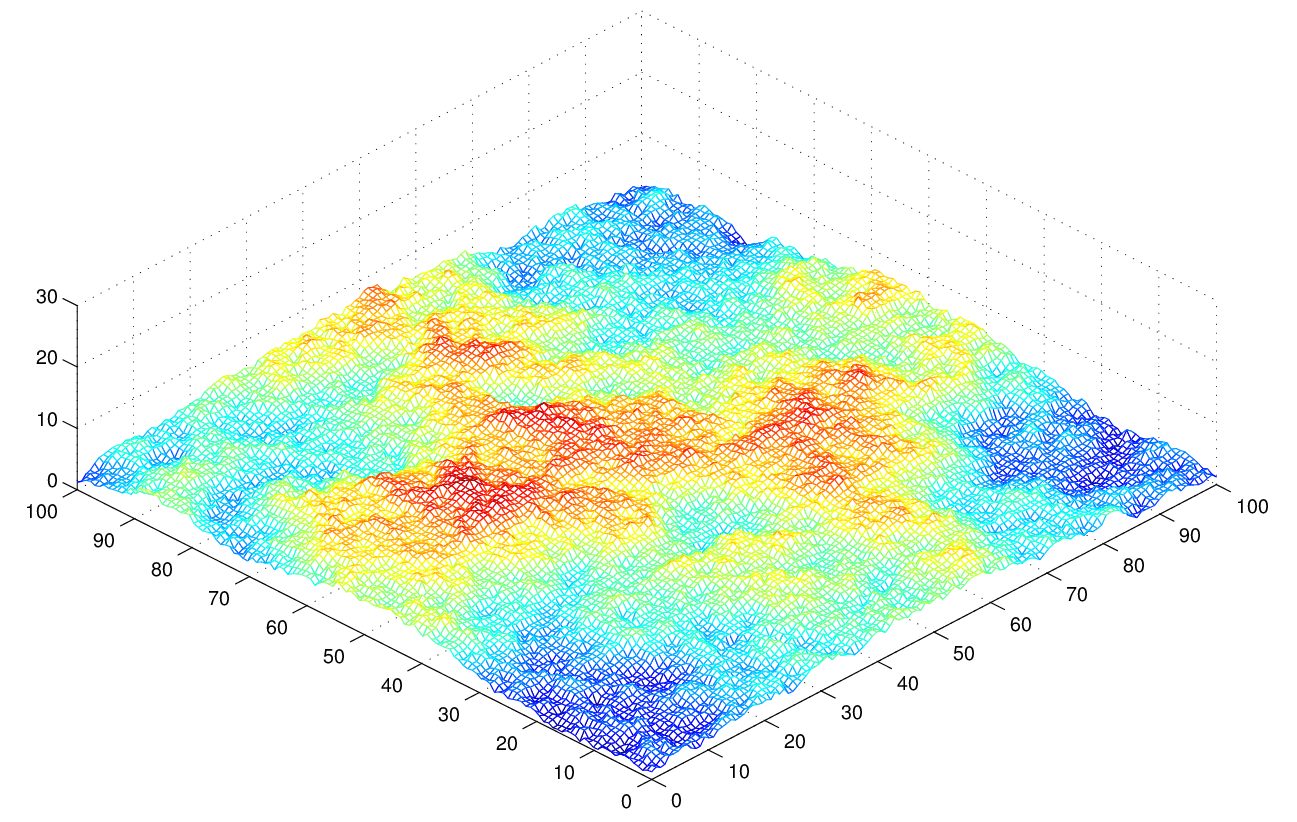}\label{fig2b}}
\subfigure[$D=2.9$]{\includegraphics[width=.32\textwidth,angle=0]{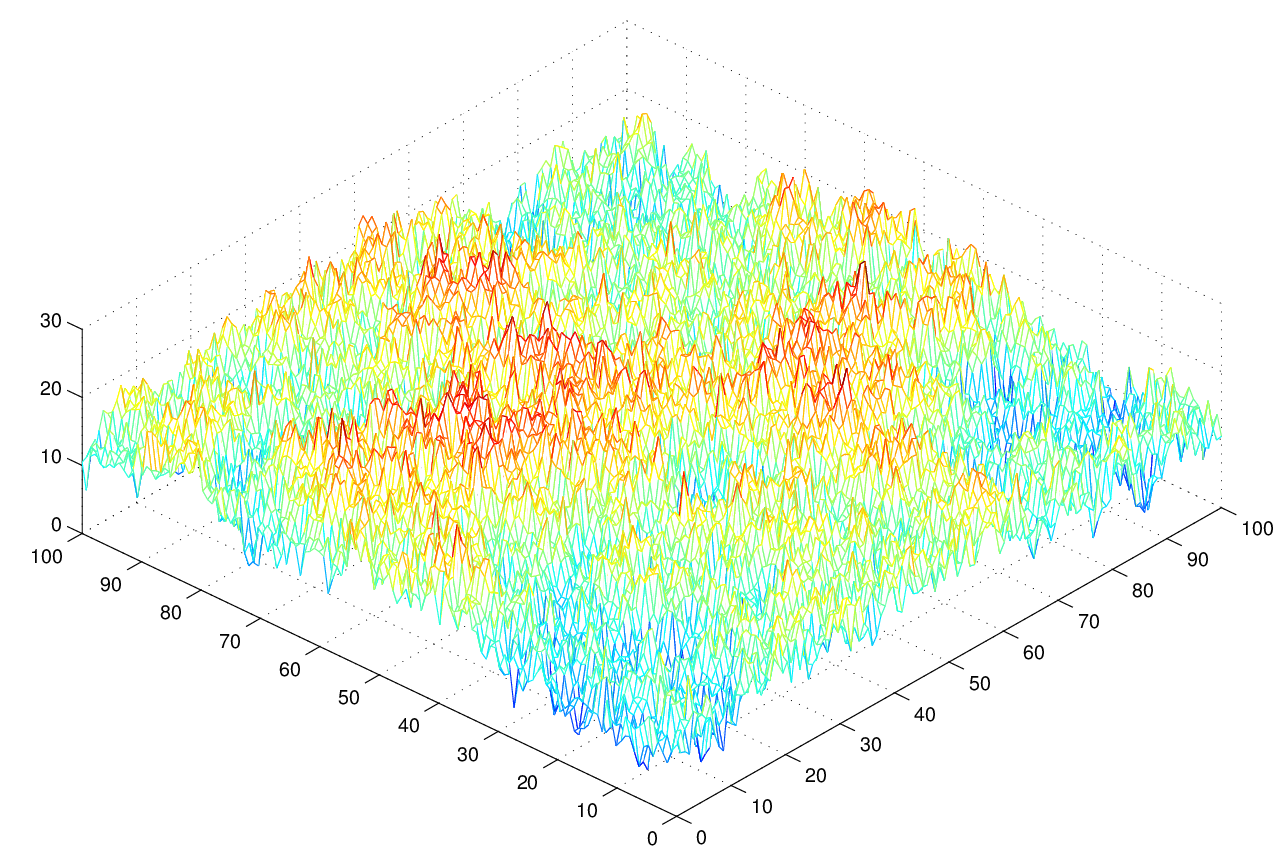}\label{fig2c}}
\caption{the effect of the fractal dimension D on the numerically
generated surfaces with $m=7$: the increase of roughness with
augmenting $D$.}\label{fig2}
\end{figure}

\begin{figure}
\centering
\subfigure[$m=5$]{\includegraphics[width=.48\textwidth,angle=0]{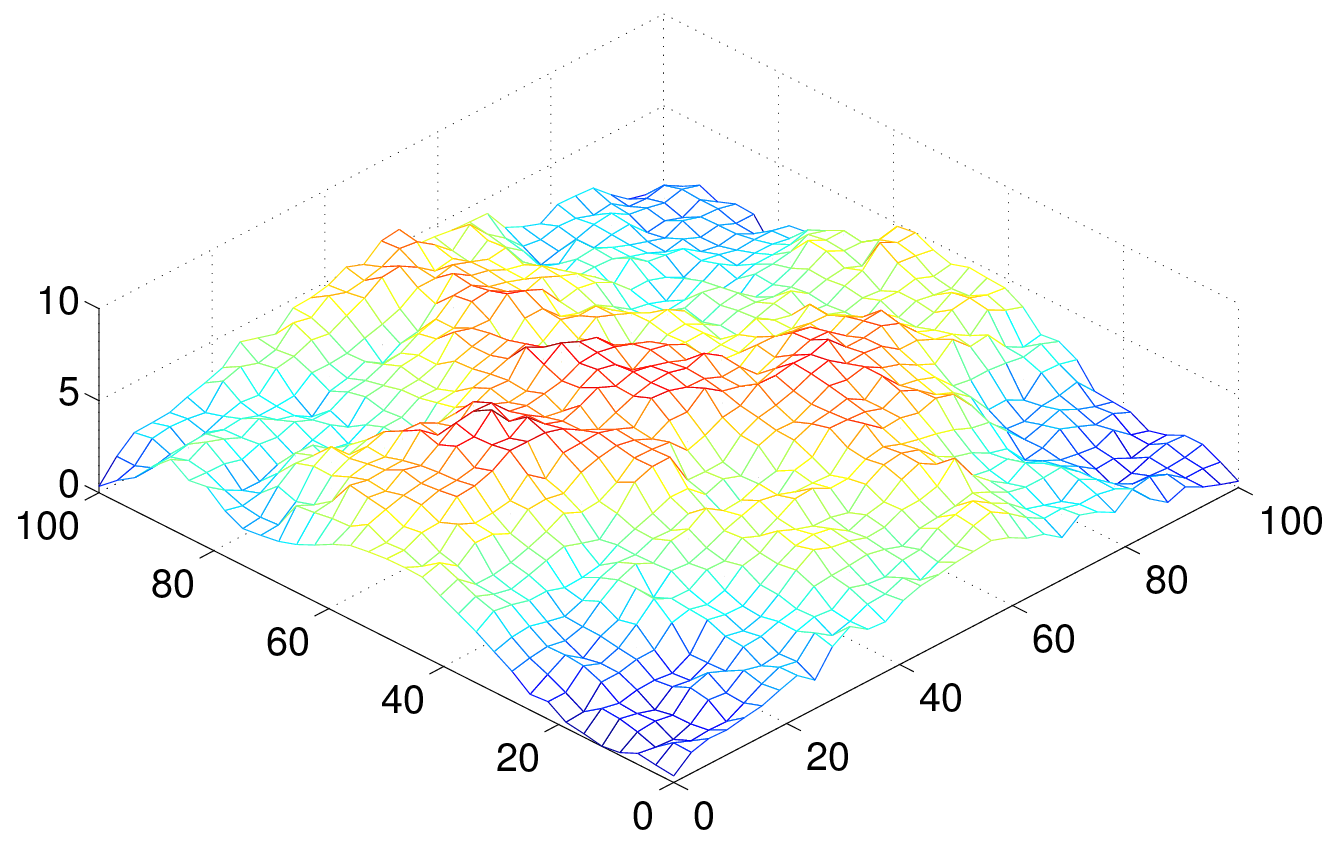}\label{fig3a}}\;
\subfigure[$m=6$]{\includegraphics[width=.48\textwidth,angle=0]{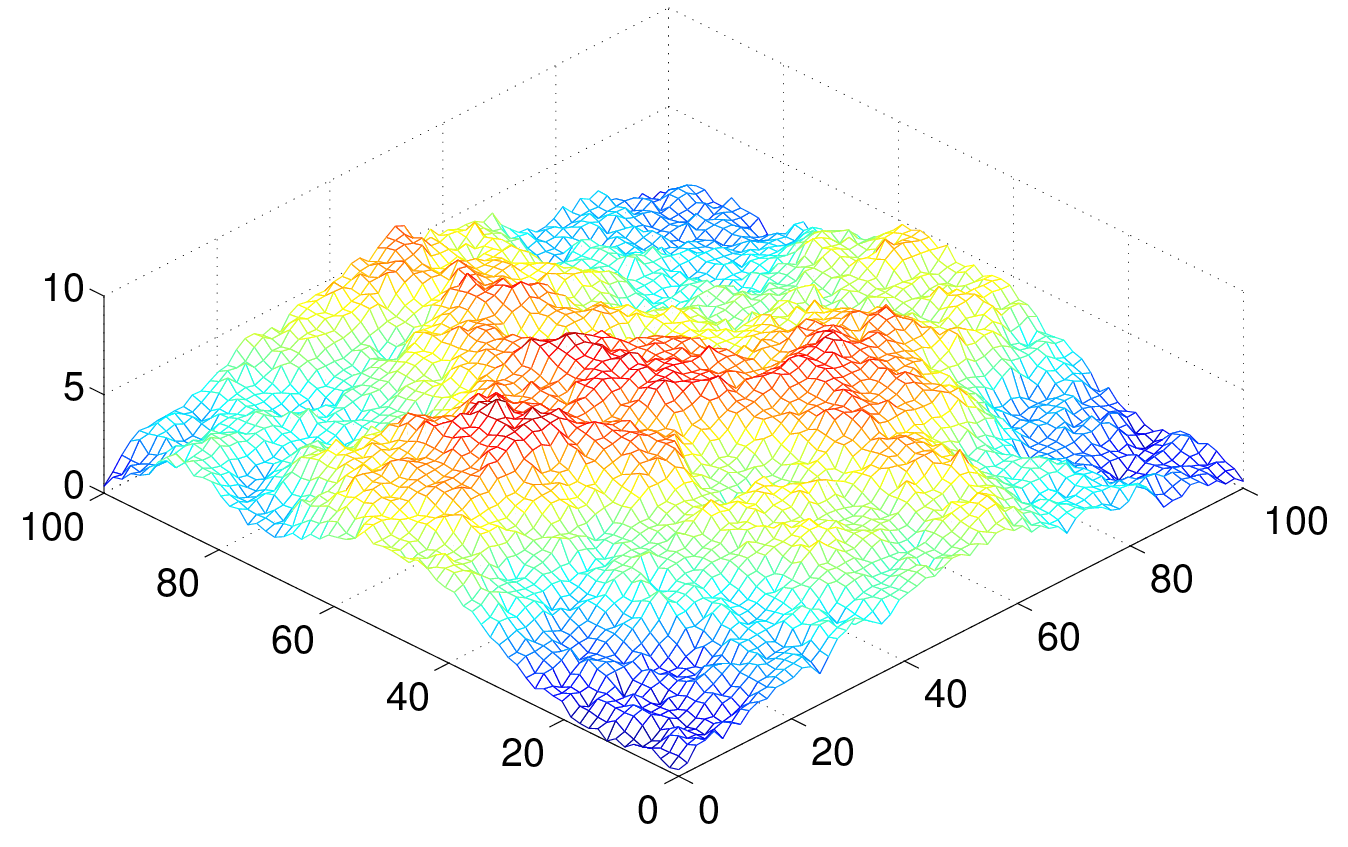}\label{fig3b}}\\
\subfigure[$m=7$]{\includegraphics[width=.48\textwidth,angle=0]{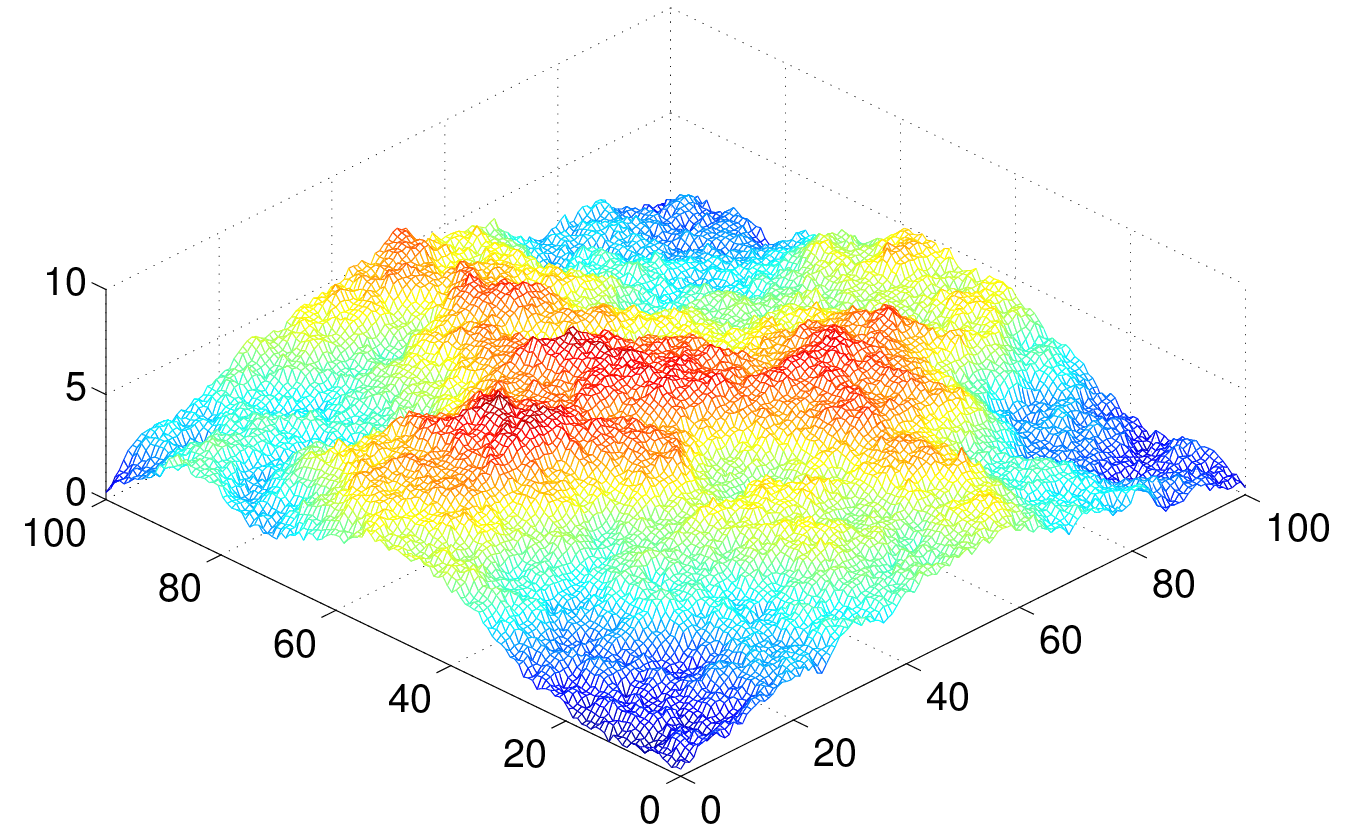}\label{fig3c}}\;
\subfigure[$m=8$]{\includegraphics[width=.48\textwidth,angle=0]{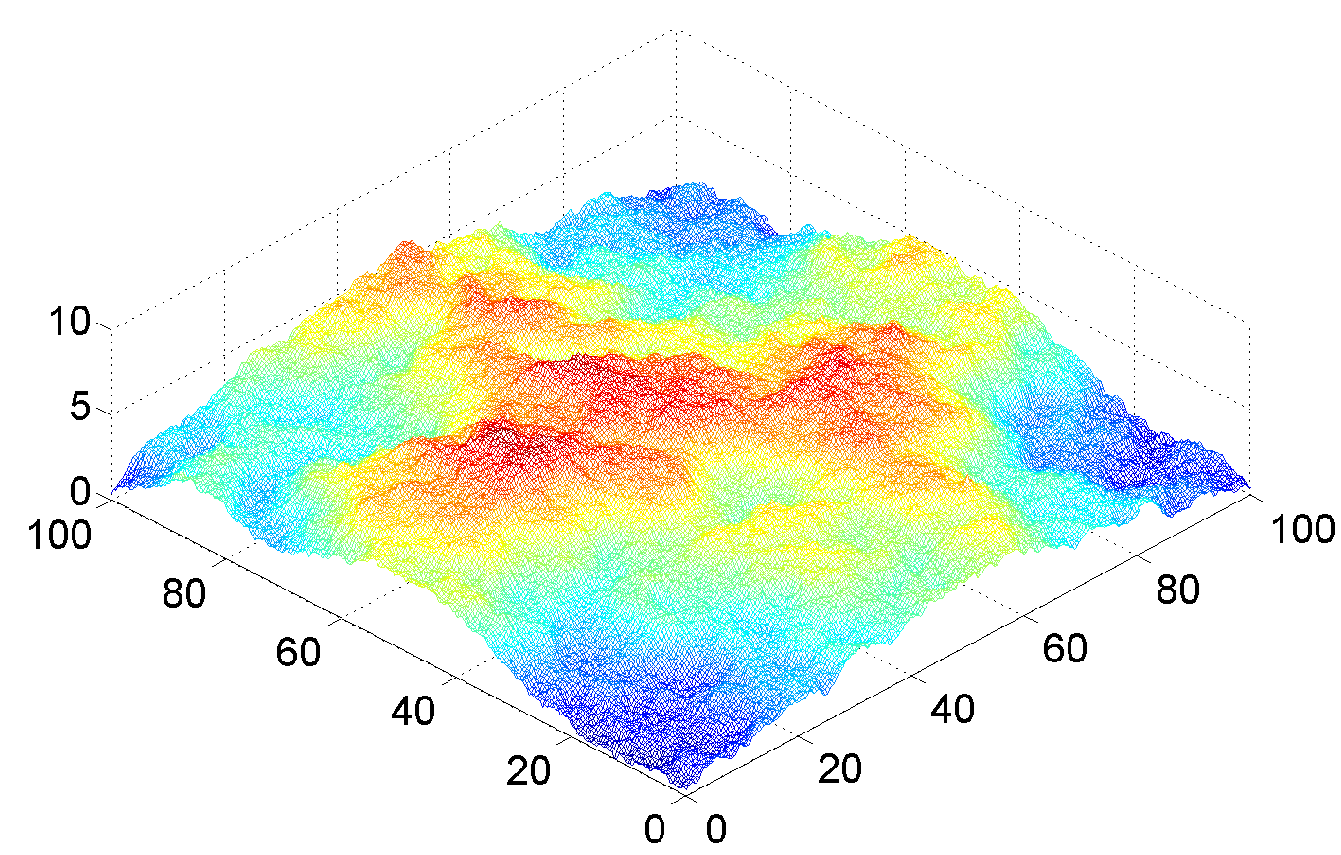}\label{fig3d}}
\caption{rough surfaces with different values of the parameter
$m$}\label{fig3}
\end{figure}

We investigate both the effect of the fractal dimension $D$, see an
example in Fig.\ref{fig2}, and the effect of the surface resolution
by varying the generation parameter $m$, see Fig.\ref{fig3}.

\begin{figure}
\centering \subfigure[The effect of
$D$]{\includegraphics[width=.42\textwidth,angle=0]{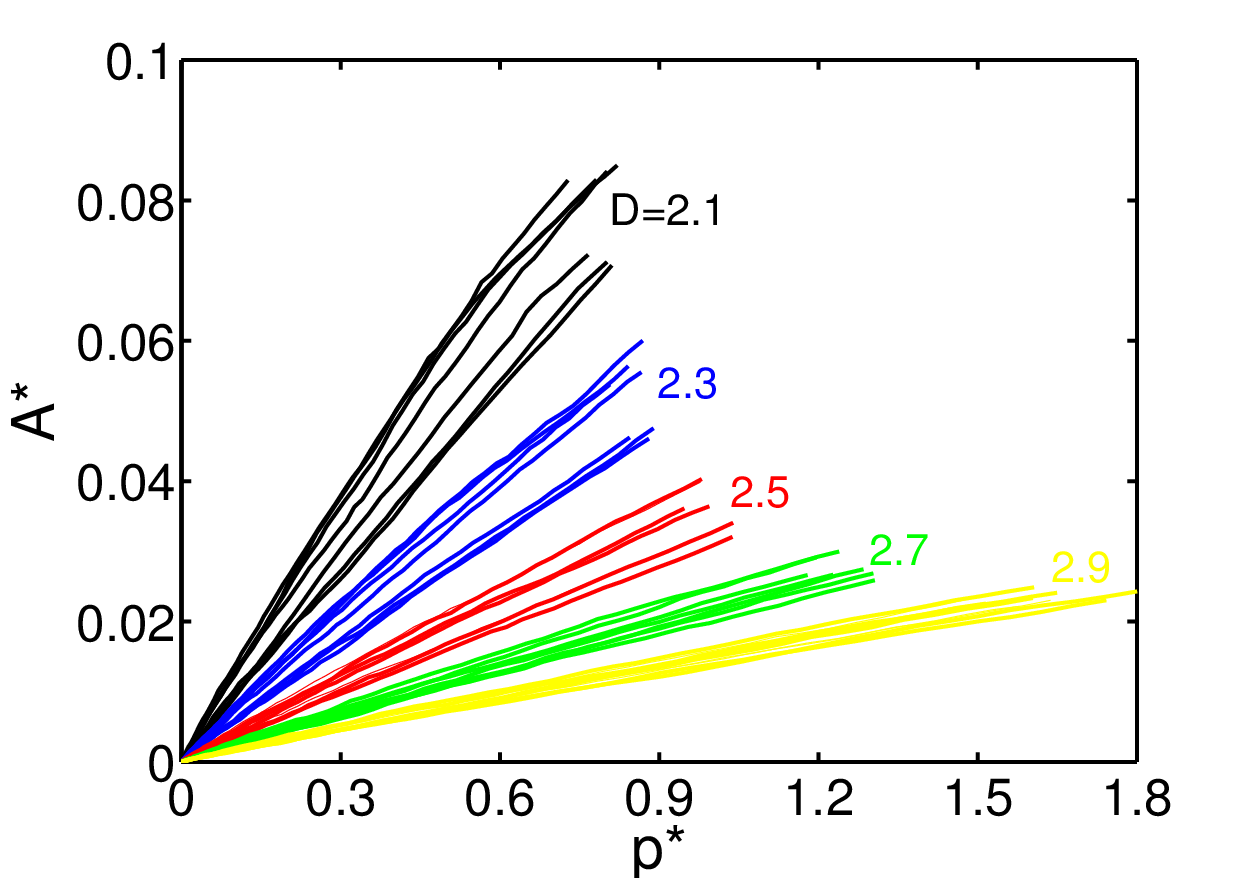}\label{fig4a}}\;
\subfigure[The effect of
$m$]{\includegraphics[width=.45\textwidth,angle=0]{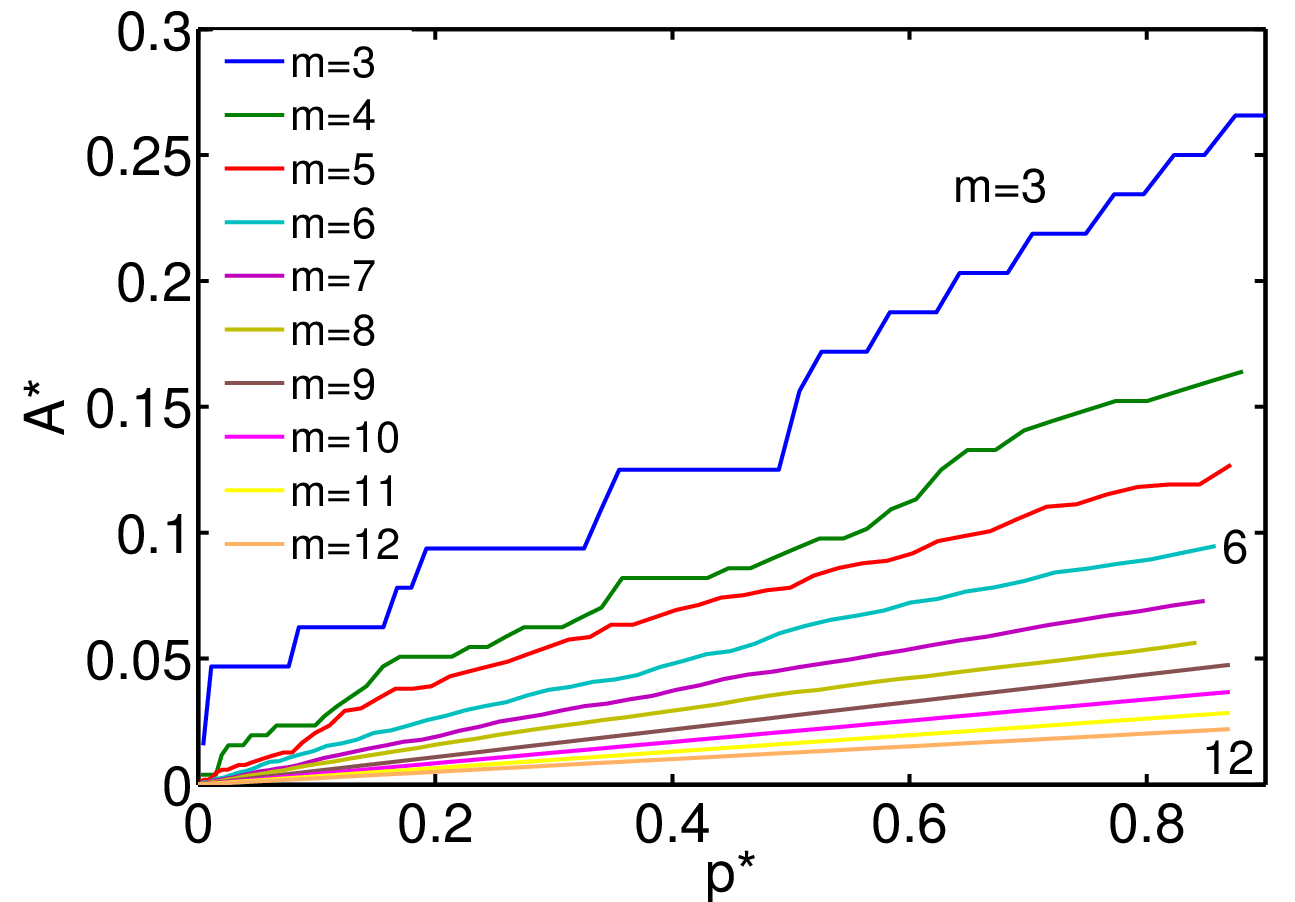}\label{fig4b}}
\caption{Real contact area fraction $A^*$ vs. dimensionless pressure
$p^*$ for numerically generated fractal surfaces. (a) The effect of
the fractal dimension $D$ (10 surfaces per set, $m=8$); (b) the
effect of the resolution parameter $m$ (surfaces with
$D=2.3$).}\label{fig4}
\end{figure}

In agreement with previous findings \cite{BCC}, the computed $A^*$
vs. $p^*$ curves depend on the surface fractal dimension $D$, see
Fig.\ref{fig4a} obtained for surfaces with $L=100$ and a resolution
determined by $m=8$, viz., 257 heights per side. Monte Carlo
simulations have been performed by considering 10 randomly generated
surfaces for each set of surface parameters. The larger the fractal
dimension $D$, the more spiky the surface with a consequently
reduced real contact area for the same given pressure. The relation
between $A^*$ and $p^*$ is also resolution-dependent due to the
lacunarity of the contact domain, see Fig.\ref{fig4b} for surfaces
with $D=2.3$, $L=100$ and different $m$. In this case we just show
the results for a single surface, since the trend is the same
regardless of the statistical variability in the surface generation.
This property was pointed out in theory \cite{MB,persson} and in
numerical simulations \cite{BCC} and it implies a vanishing real
contact area in the theoretical limit of an infinite resolution
($\delta\to 0$, or $s\to \infty$). Considering a linear
approximation for the $A^*$ vs. $p^*$ relation, we obtain a
power-law relation of the type $A^*/p^*\sim \delta^{0.37}$, which is
in fair good agreement with the predictions by Persson theory
\cite{persson} suggesting $A^*/p^*\sim \delta^{D- 2}$, i.e.
$A^*/p^*\sim\delta^{0.3}$ for the present set of surfaces with
$D=2.3$.

\begin{figure}
\centering
\includegraphics[width=.5\textwidth,angle=0]{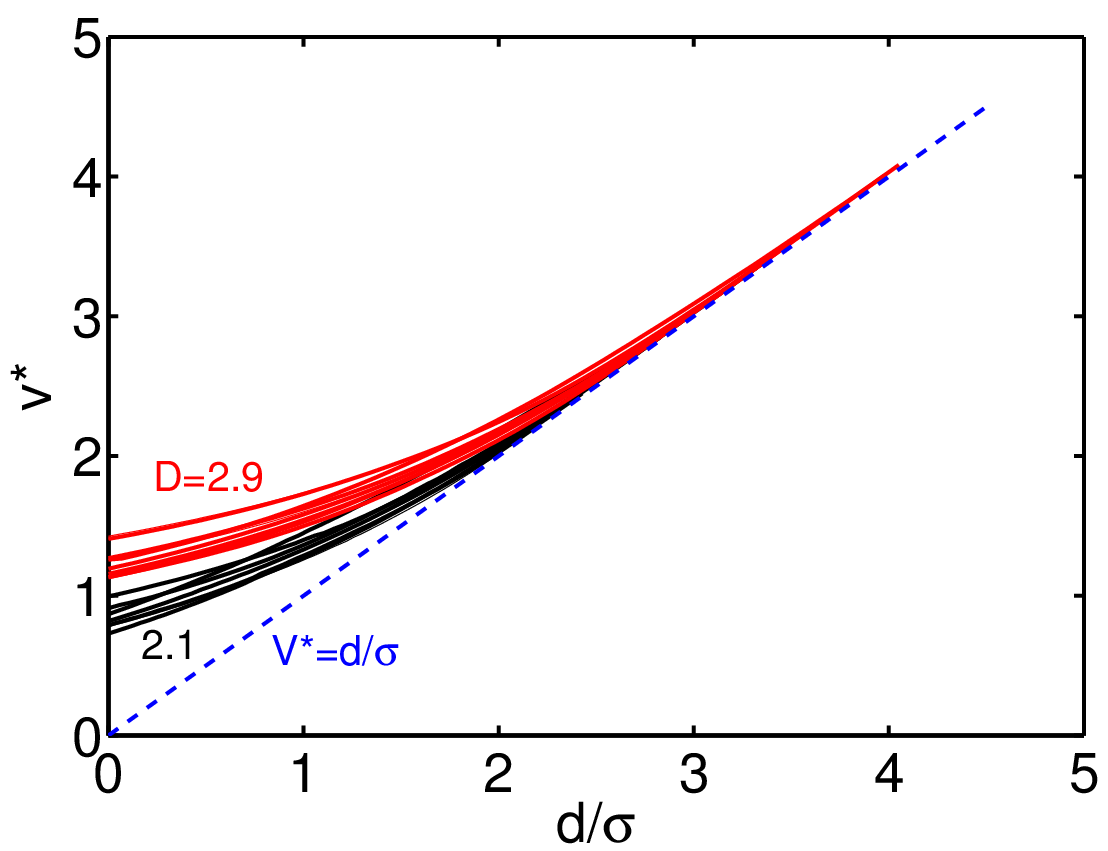}
\caption{Dependence of $V^*$ on the mean plane separation
$d/\sigma$. Note the deviation from linearity for small values of
$d/\sigma$.}\label{fig5}
\end{figure}

As far as the free volume between rough surfaces is concerned, one
might argue that this quantity should be somehow proportional to the
dimensionless mean plane separation, $d/\sigma$, where $d$ is the
separation between the half plane of the rough surface and the
elastic half space, and $\sigma$ is the r.m.s. roughness. However,
from Fig.\ref{fig5} where the contact predictions for two sets of
surfaces with $D=2.1$ (very smooth) and $D=2.9$ (very rough) are
depicted, we achieve the important result that the relation
$V^*=d/\sigma$ (depicted with a dashed blue line) holds only for
very large separations $(d/\sigma\gtrsim 3)$. For $(d/\sigma\lesssim
3)$, these two quantities cannot be confused any longer and the free
volume starts depending on the fractal dimension $D$ as well.

\begin{figure}
\centering \subfigure[$V^*$ vs.
$p^*$]{\includegraphics[width=.45\textwidth,angle=0]{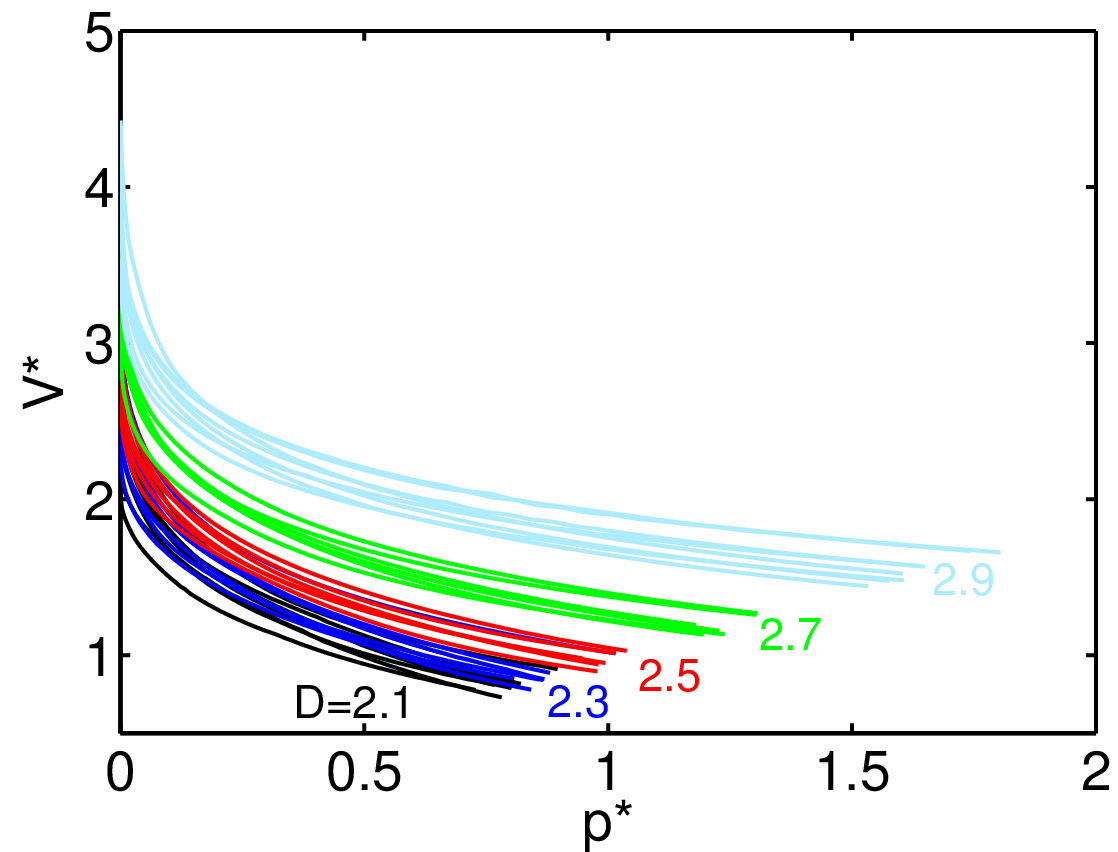}\label{fig6a}}\;
\subfigure[$V^*$ vs.
$A^*$]{\includegraphics[width=.45\textwidth,angle=0]{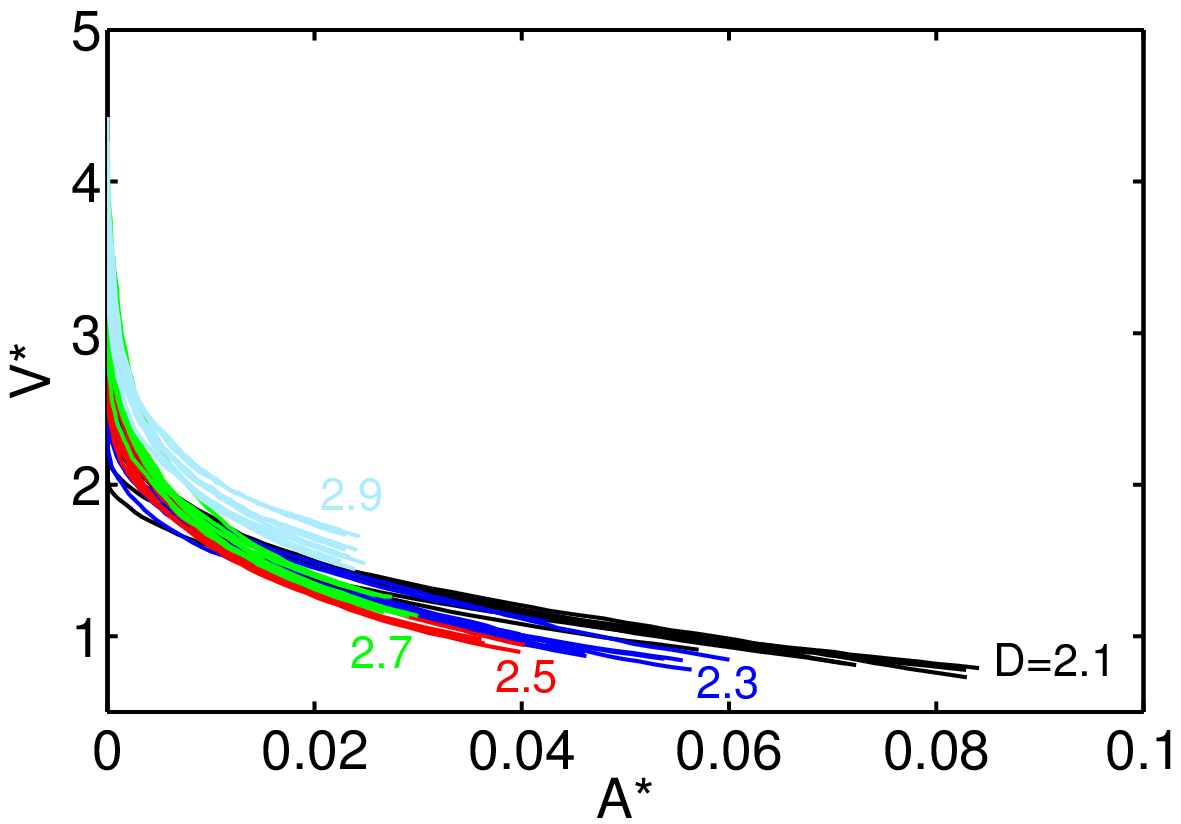}\label{fig6b}}
\caption{Dimensionless volume $V^*$ vs. dimensionless nominal
contact pressure $p^*$ and real contact area fraction $A^*$, for
various fractal dimensions $D$ (10 surfaces per set,
$m=8$).}\label{fig6}
\end{figure}

Investigating the relation between $V^*$ and the other contact
quantities, namely $A^*$ and $p^*$, by varying $D$, we obtain the
diagrams shown in Fig.\ref{fig6}. A decay of the free volume by
increasing the dimensionless contact pressure or the dimensionless
real contact area is observed. The relation between $V^*$ and $p^*$
is significantly affected by $D$ (Fig.\ref{fig6a}). On the other
hand, as a notable result, the relation between $V^*$ and $A^*$
seems to be almost independent of the fractal dimension, since all
the curves lie in a relatively narrow band (Fig.\ref{fig6b}).

\begin{figure}
\centering \subfigure[$V^*$ vs.
$p^*$]{\includegraphics[width=.42\textwidth,angle=0]{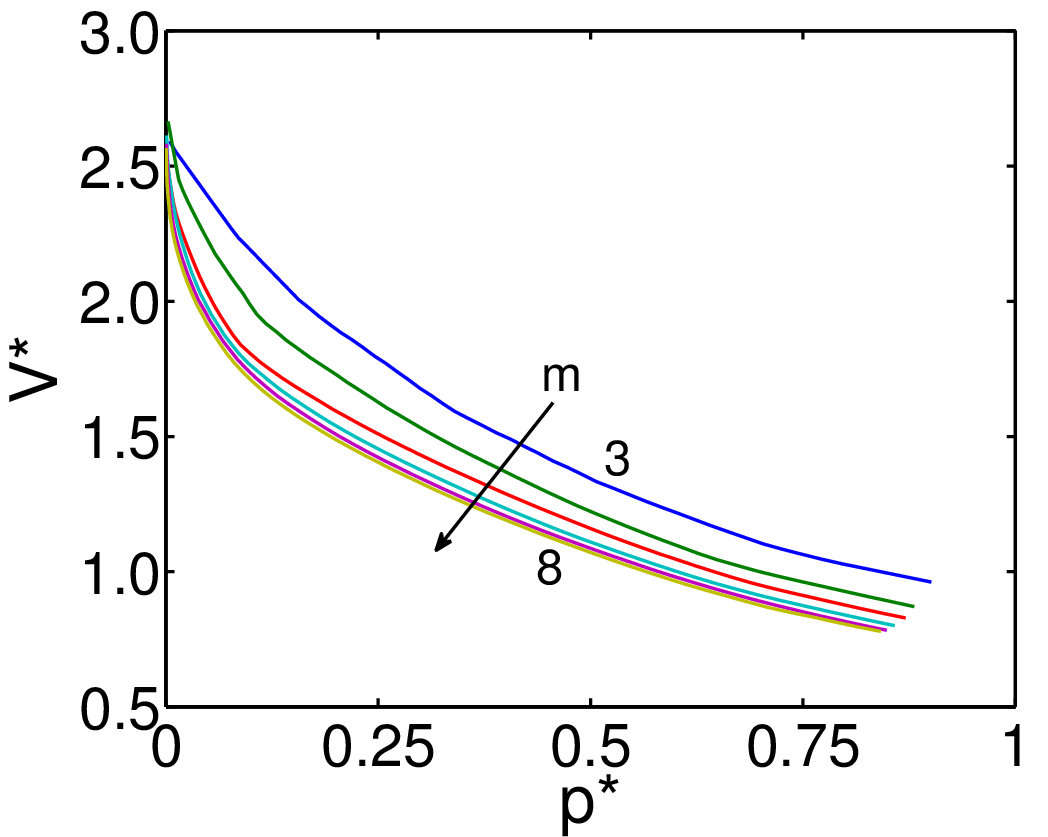}\label{fig7a}}\;
\subfigure[$V^*$ vs.
$A^*$]{\includegraphics[width=.40\textwidth,angle=0]{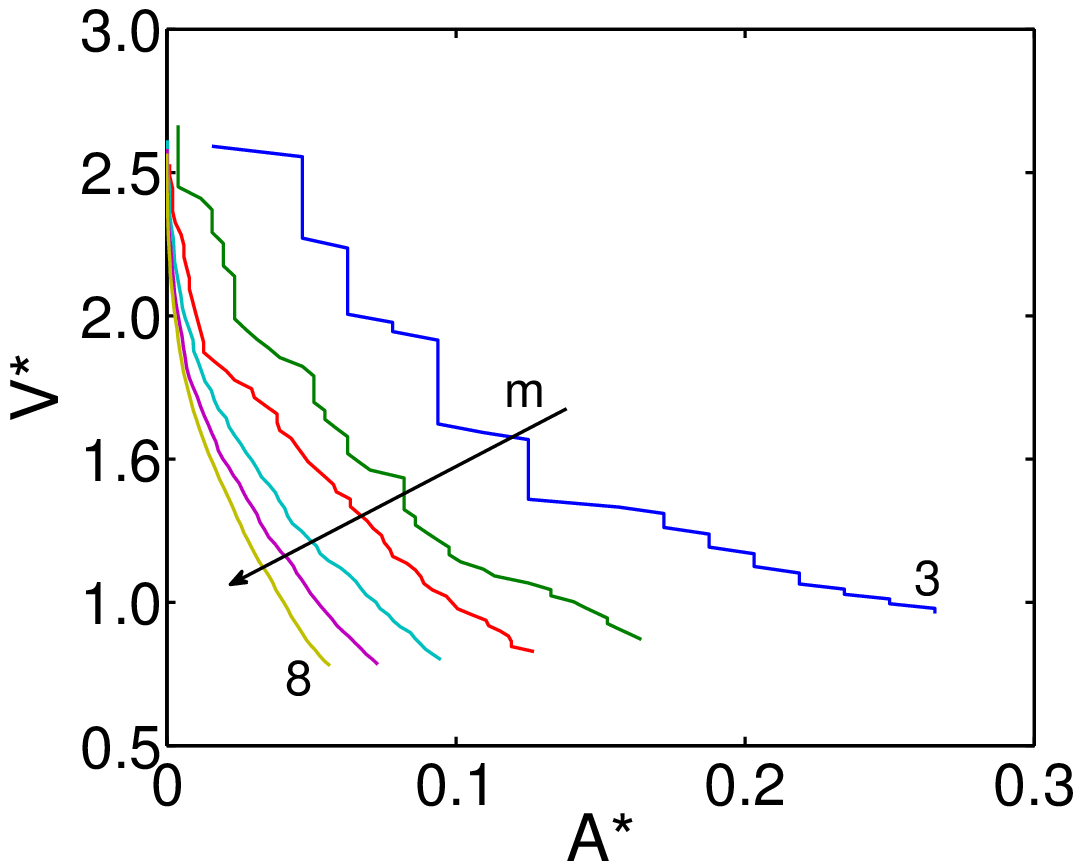}\label{fig7b}}
\caption{Dimensionless volume $V^*$ vs. dimensionless nominal
contact pressure $p^*$ and real contact area fraction $A^*$, for
various values of the surface resolution parameter $m$ ad
$D=2.3$.}\label{fig7}
\end{figure}

To examine the role played by the surface resolution, we now
consider a single surface with $D=2.3$ and we change the resolution
parameter $m$. The trends shown in Fig.\ref{fig7} pinpoint a
convergence of the relation $V^*$ vs. $p^*$ by increasing $m$. On
the other hand, the relation $V^*$ vs. $A^*$ is strongly resolution
dependent.

\subsection{Fractal properties of the free volume domains}

\begin{figure}
\centering \subfigure[Undeformed surface
$(A^*=0)$]{\includegraphics[width=.42\textwidth,angle=0]{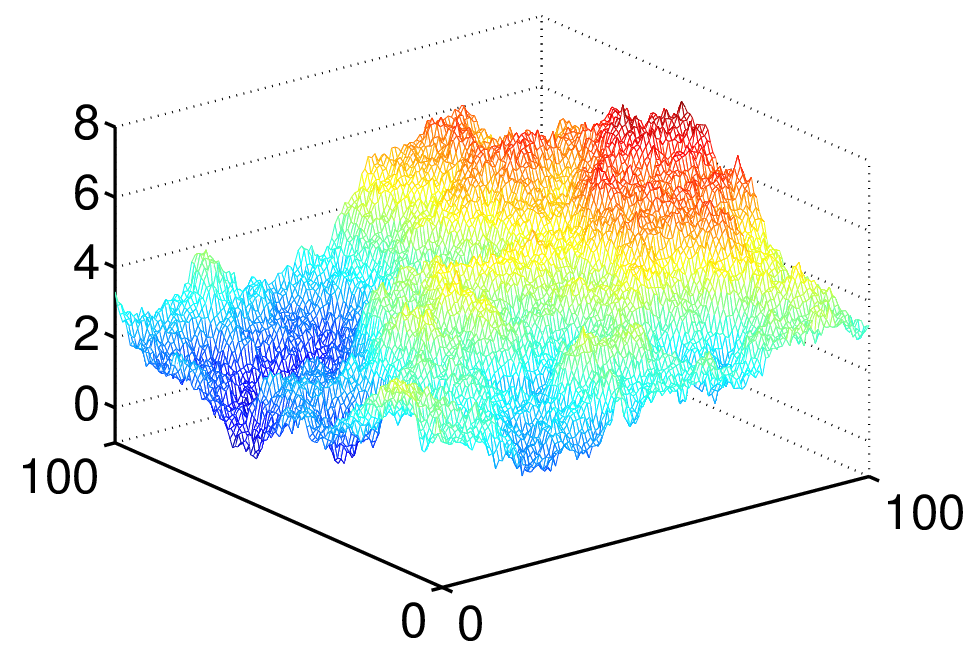}\label{fig8a}}
\subfigure[Deformed surface $(A^*\sim
0.1)$]{\includegraphics[width=.42\textwidth,angle=0]{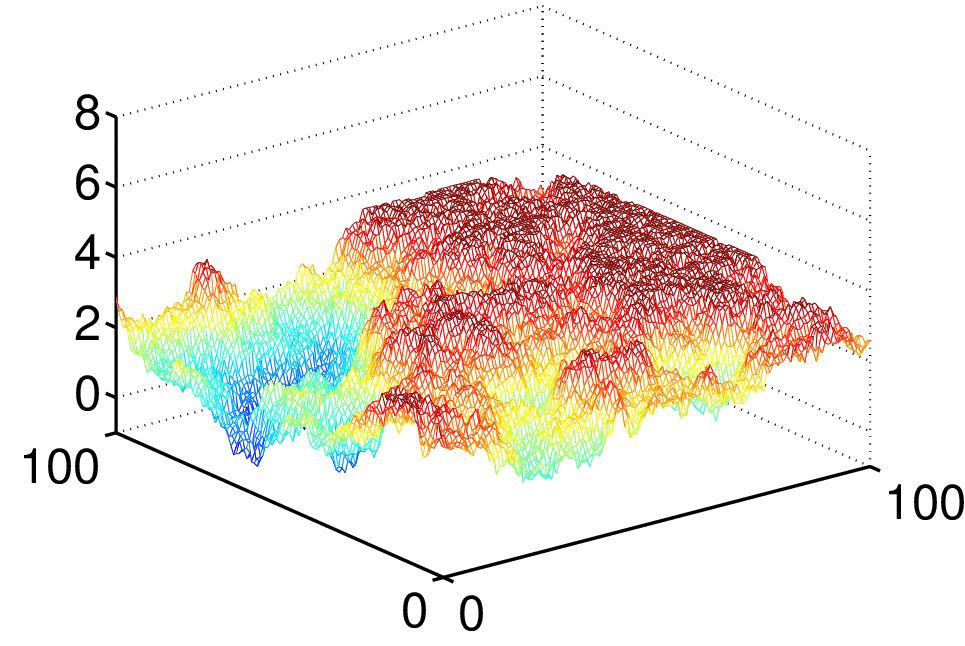}\label{fig8b}}
\caption{Original undeformed surface and its deformed shape
corresponding to $A^*\sim 0.1$.}\label{fig8}
\end{figure}

From the results of numerical simulations it is possible to
visualize the deformed configuration of the rough surface in contact
with the half-plane for each imposed displacement, see the
undeformed shape of a surface with $D=2.3$, $L=100$ and $m=7$ in
Fig.\ref{fig8a} and its deformed shape corresponding to a real
contact area fraction $A^*\sim 0.1$ in Fig.\ref{fig8b}.

\begin{figure}
\centering
\includegraphics[width=.45\textwidth,angle=0]{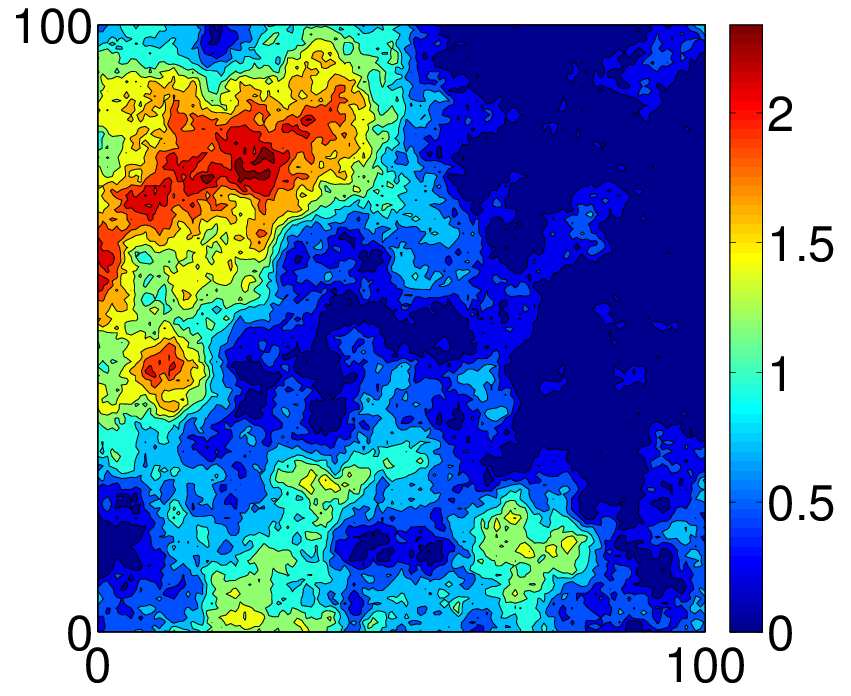}
\caption{Contour plot of the free volumes $v_{i,j}$ related to
Fig.\ref{fig8b}. Red denotes deep valleys with larger free volumes
not in contact, blue denotes asperities in contact with zero free
volume remaining (see the online version for colours).}\label{fig9}
\end{figure}

The spatial distribution of the real contact area and the amount of
the free volume $v_{i,j}$ at each grid point can also be visualized.
Here the total free volume is $V=\sum_{i,j} v_{i,j}$, where $i,j$
are indices running over all the boundary elements. Due to
roughness, the asperities, which are the maxima of the 3D surface,
come into contact at isolated points and then progressively merge
together by forming wider contact regions with zero free volume.
Other regions of the surface present free volumes $v_{i,j}$ whose
size depends on the amplitude of the valleys. A contour plot in
Fig.\ref{fig9} corresponding to the deformed configuration in
Fig.\ref{fig8b} displays the areas with different values of
$v_{i,j}$. The dark red color denotes the contours having the
highest values of $v_{i,j}$, whereas the deepest blue represents the
contour for $v_{i,j}=0$. The morphological properties of the surface
valleys, usually neglected in contact problems where the real
contact area is the primary quantity of interest, are indeed
relevant for the spatial distribution of the free volume. Existing
standards suggest to use the indices $Sbi$ and $Svi$ to quantify the
free volume properties of surfaces. In particular, the valley fluid
retention index, $Svi$, computed from the bearing area curve as the
volume comprised between the undeformed surface and the plane
leaving only $20\%$ of the heights below it, and divided by the
product $\sigma A_n$, is certainly a useful indicator to distinguish
between surfaces with very large or small valleys. However, these
roughness parameters computed from the original undeformed geometry
do not account for the effect of asperity deformations occurring
during contact.

\begin{figure}
\centering
\includegraphics[width=.65\textwidth,angle=0]{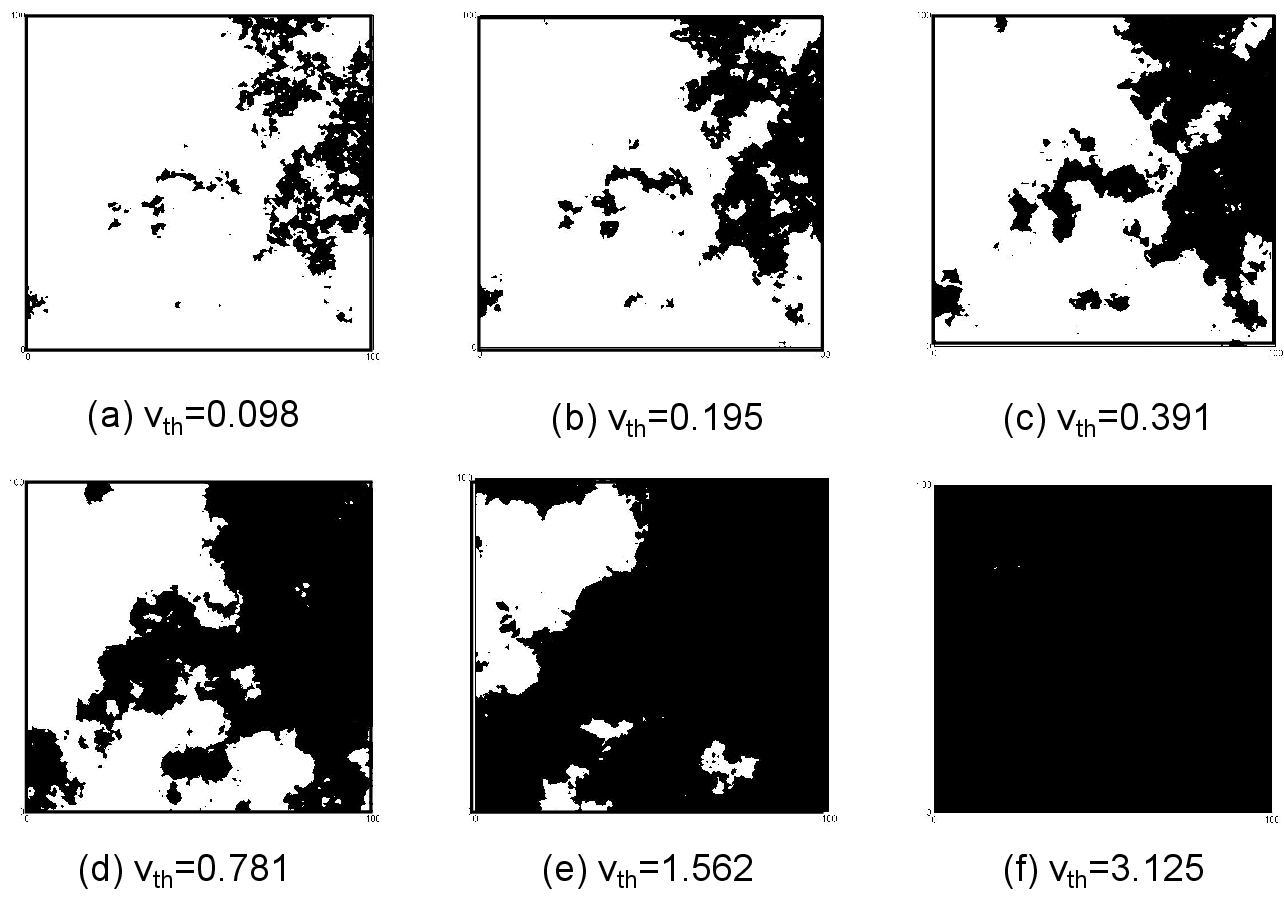}
\caption{Free volume domains (black areas) $\mathbb{D}(v_{i,j}\leq
v_{th})$ corresponding to Fig.\ref{fig9} for different free volume
thresholds $v_{th}$.}\label{fig10}
\end{figure}

A deeper insight into the morphological properties of the spatial
distribution of the free volume can be made by examining the contour
levels corresponding to different volume thresholds, $v_{th}$, as
shown in Fig.\ref{fig10}. In these contours, the black area denotes
domains $\mathbb{D}$ where $v_{i,j}\leq v_{th}$. Therefore, the dark
islands for the limit case of $v_{th}=0$ correspond to the real
contact area domain. Selecting $v_{th}$ larger than the maximum
value of the volume of the deepest valley, $v_{th}=\max(v_{i,j})$,
the picture becomes entirely black since all the elementary areas of
the grid have $v_{i,j}\leq v_{th}$. This second limit situation
corresponds to the Euclidean domain of the nominal contact area.

It has to be remarked that the contour plots in Fig.\ref{fig10},
corresponding to the same contact pressure and contact area,
dynamically change during contact. At first contact, the real
contact area $A^*$ is zero and the free volume $V^*$ is maximum. By
increasing the contact pressure, the domain of the real contact area
increases until it reaches $A^*=1$ and full contact takes place.
Conversely, the free volume domain progressively shrinks down to
zero in the same limit. It has to be pointed out that the observed
scaling of the real contact area is in agreement with former
numerical investigations, see \cite{BCC,YAM}.

For each contour plot in Fig.\ref{fig10}, the topological properties
of the free volume domains $\mathbb{D}(v_{i,j}\leq v_{th})$ can be
investigated according to the box counting method. For $A^*=0$
(undeformed rough surface), the free volume domains are expected to
be self-affine as a consequence of the self-affinity of the parent
surface \cite{bigerelle1}. For a value $0<A^*\le 1$, on the other
hand, the topological properties of the free volume domains have to
be correlated with those of the deformed surface whose heights have
been modified by elastic deformation.

For each box of lateral size $r$, the number $N$ of boxes containing
at least one black grid point are counted. This operation is
repeated by varying $r$ from $1$ up to $2^m$ lateral size divisions,
with a geometric progression of 2. The cumulative number $N(r)$ is
plotted vs. $r$ in a bi-logarithmic diagram and the local fractal
dimension $\mathcal{D}$ of the volume domain can be finally obtained
by differentiating $\log(N)$ w.r.t. $\log(r)$.

\begin{figure}
\centering \subfigure[Box-counting
plot]{\includegraphics[width=.42\textwidth,angle=0]{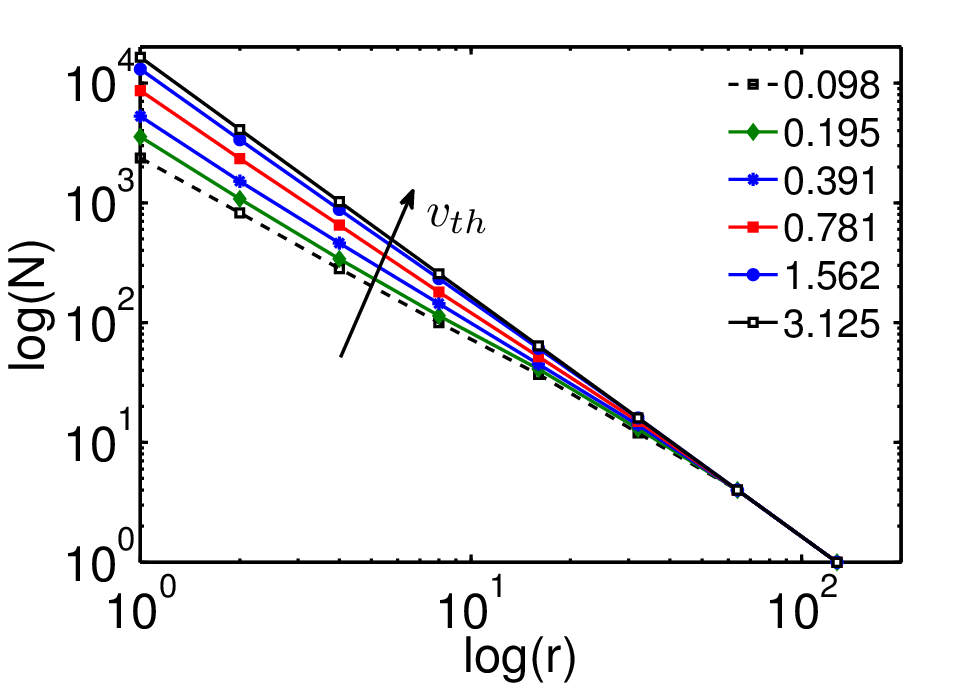}\label{fig11a}}\;
\subfigure[Local fractal
dimension]{\includegraphics[width=.42\textwidth,angle=0]{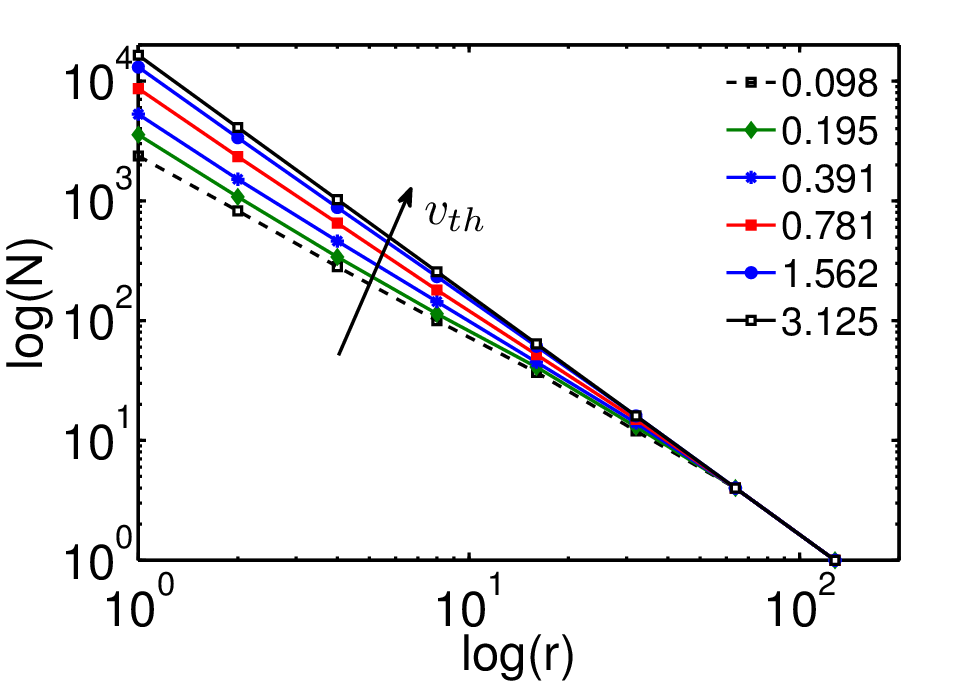}\label{fig11b}}
\caption{Fractal characterization of the free volume domains
$\mathbb{D}(v_{i,j}\leq v_{th})$ shown in Fig.\ref{fig10}.
Fig.\ref{fig8b} shows how the local fractal dimension $\mathcal{D}$
varies as a function of $r$ and $v_{th}$.}\label{fig11}
\end{figure}

By performing this analysis for the domains $\mathbb{D}$ in
Fig.\ref{fig10}, we obtain the diagram in Fig.\ref{fig11a}. The
curves have a trend close to a straight line in this bi-logarithmic
plot, which suggests a power-law scaling typical of fractals.

The local fractal dimension is shown in Fig.\ref{fig11b} and it is
found to be dependent on $r$. In the limit case corresponding to
$v_{th}=0$, the fractal dimension of the corresponding free volume
domain is equal to that of the real contact area, which is less than
2 due to the lacunarity of the contact domain. According to the
results in \cite{BCC01}, the fractal dimension of the contact area
is an increasing function of the applied pressure but it is a
decreasing function of $D$. In the other limit scenario of
$v_{th}=\max(v_{i,j})$, the fractal dimension is equal to $2$, i.e.,
it is equal to that of an Euclidean smooth surface. These limit
values represent the \emph{bounds} to the fractal dimension of the
free volume contours $\mathbb{D}(v_{i,j}\leq v_{th})$ by varying
$v_{th}$.

\section{Theoretical considerations}

Apart from the direct post-processing of the numerical results, a
theoretical formula for the computation of $V^*$ could be proposed
by following the principle inspiring the \emph{bearing area curve}
\cite{AF33}, which is also the base for the definition of the valley
fluid retention index $Svi$ \cite{standard1,standard2}. According to
this reasoning, the free volume for a given indentation $\Delta$
(related to the elevation $h$ of the contacting plane from the
average plane of the surface heights determined in the undeformed
configuration) can be estimated via the integral of the vertical gap
$(h-z)$ times the distribution density function $\Phi(z)$ of the
surface heights, multiplied by the real area not in contact:
\begin{equation}\label{bearing}
V^*=\dfrac{A_n-A_r}{\sigma
A_n}\int_{-\infty}^h(h-z)\Phi(z)\mbox{d}z=\dfrac{1-A^*}{\sigma}\int_{-\infty}^h(h-z)\Phi(z)\mbox{d}z,
\end{equation}
where the integration limits span over the whole range of heights
having a positive gap, i.e., from $-\infty$ up to $h$.

It is remarkable to note that, although this approach provides a
decay of $V^*$ by increasing $A^*$, this would be simply linear if
the distribution $\Phi(z)$ is assumed to be a Gaussian distribution,
as often put forward in the literature (see \cite{ZBP04} for a
review of micromechanical contact theories making this assumptions).
This is not exactly the case observed in numerical experiments, see
Fig.\ref{fig3b}. Moreover, the predictions of Eq.\eqref{bearing} are
independent of $D$ and $m$, unless an additional relation $A^*(D,m)$
is invoked. Hence, according to Eq.\eqref{bearing}, the dependency
of $V^*$ on the fractality of surfaces and on their resolution would
be an indirect consequence of the dependency of $A^*$ on $D$ and
$m$, as also takes place in the model based on Persson's theory
\cite{bottiglione}.

An alternative and more accurate path to evaluate the free volume is
to exploit the fractal properties of the corresponding domains
$\mathbb{D}$ analyzed for instance in Fig.\ref{fig11}. This approach
leads to a formula for the estimation of $V^*$ independent from
Eq.\eqref{bearing}. This has the advantage of being able to explain
the dependencies of $V^*$ on the fractal dimension and on the
surface resolution solely from the statistical distribution of the
free volume, without invoking the scaling properties of other
contact predictions like the real contact area.

The main consequence of the plot in Fig.\ref{fig11a} is that the
cumulative number of boxes with a volume $v_{i,j}<v_{th}$ has a
power-law scaling with respect to the box size $r$:
\begin{equation}\label{eq1}
N(r)=\left(R/r\right)^{\mathcal{D}},
\end{equation}
where $R$ is a free parameter to be determined from data. Equation
\eqref{eq1} can be recast in terms of a lateral size $l$ and of the
sampling interval $\delta$, with again $l$ as a free parameter:
\begin{equation}\label{eq2}
N(\delta)=\left(l/\delta\right)^{\mathcal{D}}.
\end{equation}
Here, $\mathcal{D}$ should be considered as dependent on $r$ and
$v_{th}$, according to the trends shown in Fig.\ref{fig11b}.

For a given surface resolution, the total free volume $V^*$ can be
computed according to the following integral:
\begin{equation}\label{int}
V^*=\dfrac{1}{\sigma A_n}\int_{0}^{v_{th,\max}}n\,\mbox{d}v_{th},
\end{equation}
where $n$ represents the number of boxes with a free volume
comprised in the range $v_{th}<v_{i,j}<v_{th}+\mbox{d}v_{th}$.

A closed-form expression to $n$ appears to be difficult to be
derived, since this is a nonlinear function of $v_{th}$. However, it
is possible to proceed with a numerical integration of
Eq.\eqref{int} by suitably partioning the limit of integration
$v_{th,\max}$ in $K$ intervals $(v_{th,k},v_{th,k+1})$, where
$k=1,\dots,K$, and replacing the integration by a discrete sum. The
number of boxes $n$ can be approximated by the finite difference
between $N(v_{th,k+1})$ and $N(v_{th,k})$:
\begin{equation}\label{eq3}
n_k\cong\left(l/\delta\right)^{\mathcal{D}_{k+1}}-\left(l/\delta\right)^{\mathcal{D}_k},
\end{equation}
where $\mathcal{D}_k$ and $\mathcal{D}_{k+1}$ denote the fractal
dimensions of the free volume domains corresponding to $v_{th,k}$
and $v_{th,k+1}$, respectively.

The free volume of the surface comprised in the range
$v_{th,k}<v_{i,j}<v_{th,k+1}$ is therefore:
\begin{equation}\label{eq4}
\Delta V_k=n_k\Delta v_{th},
\end{equation}
where $\Delta v_{th}=v_{th,k+1}-v_{th,k}$. The total volume $V^*$
can be finally determined as
\begin{equation}\label{eq5}
V^*=\dfrac{1}{\sigma A_n}\sum_{k=1}^{K} \Delta V_k
\end{equation}

Equation \eqref{int} can be used to interpret the numerical trends
observed in Section 3.1 via the dependency of the fuction
$n(v_{i,j})$ on $D$ and $\delta$. However, instead of examining
$n(v_{i,j})$, it is more elegant and general to investigate the
probability density function $f(v)$ that can be deduced from the
numerically determined histograms $n(v_{i,j})$. The values of the
probability density function $f(v_{i,j})$ are simply determined from
$n(v_{i,j})$ as $n(v_{i,j})/n_{\text{tot}}$ and imposing the
condition $\int_0^{v_{th,\max}}f(v)\mbox{d}v=1$.

As an example, let us focus on the contact configurations
corresponding to a maximum indentation of the half-plane such that
its final position coincides with the average plane of the
originally undeformed rough surface. The probability density
function $f(v)$ is determined from BEM results corresponding to the
fractal surfaces with different $D$ and $m$ whose contact response
has been shown in Section 3.1.

Investigating the effect of the surface fractal dimension on the
free volume, we observe that $f(v)$ is a decreasing function of $v$
with an approximately linear decay, see Fig.\ref{fig11a}. By
decreasing $D$, $v_{th,\max}$ is reduced and the probability density
function becomes steeper and steeper. Interestingly, the point
$f(v=1)$ is almost independent of $D$ and acts as a fulcrum about
which the function $f$ rotates. Clearly, this effect has a double
implication on the computation of $V^*$ according to Eq.\eqref{int}:
the integration interval defined by $v_{th,\max}$ is reduced by
decreasing $D$ and the probability density function values increase
for $v\to 0$. As a consequence, the value of the free volume $V^*$
at the maximum contact area (or at the maximum pressure) is
diminishing by reducing $D$, thus explaining the trends highlighted
in Fig.\ref{fig6}.

\begin{figure}
\centering \subfigure[The effect of $D$
($m=7$)]{\includegraphics[width=.45\textwidth,angle=0]{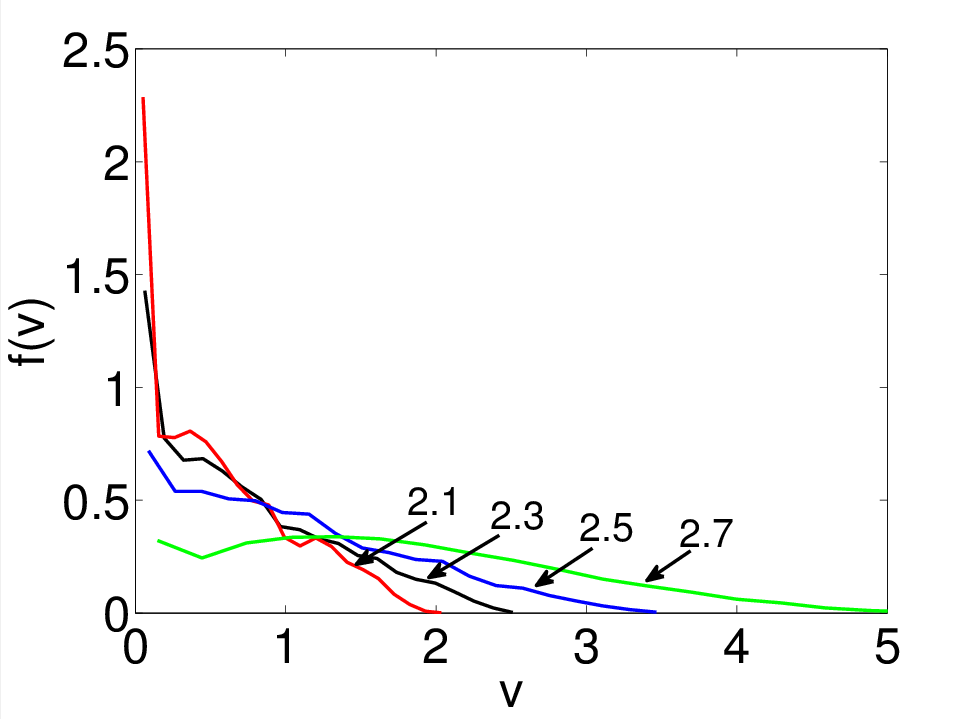}\label{fig12a}}\;
\subfigure[The effect of $m$
($D=2.3$)]{\includegraphics[width=.47\textwidth,angle=0]{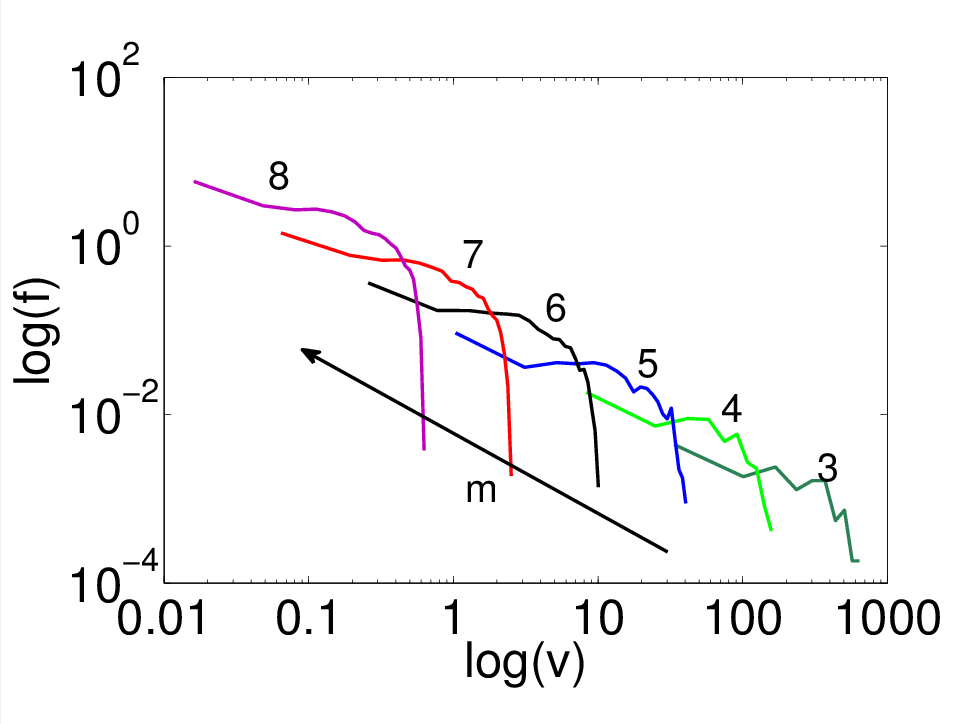}\label{fig12b}}
\caption{The effect of the surface fractal dimension $D$ and of the
surface resolution related to the generation parameter $m$ on the
probability distribution function $f(v)$ of the free volumes
$v$.}\label{fig12}
\end{figure}

As far as the effect of the surface resolution is concerned, we have
to examine the plot of the probability distribution functions for
surfaces having $D=2.3$ and different $m$ ranging from 3 to 8, see
Fig.\ref{fig12b}. The function $f(v)$ is again a decreasing function
of $v$ with an approximately linear decay as in Fig.\ref{fig12a}.
However, changing the parameter $m$ leads to a significant shift in
the probability density function values and in $v_{th,\max}$, with
variations of up to 4 orders of magnitude. Therefore, a
bilogarithmic scale has been adopted in Fig.\ref{fig12b} for
visualization purposes. By increasing $m$, the increased value of
$f$ for small values of $v$ is counterbalanced by the much smaller
value of $v_{th,\max}$. The net result stemming from the integration
$\eqref{int}$ is a decay in the total free volume $V^*$ by
increasing $m$, which provides the explanation to the trends
observed in Fig.\ref{fig7}.

An explanation to the shift of the curves in Fig.\ref{fig12b} by
varying $m$ can be provided based on the theoretical findings in
\cite{ref}. By increasing $m$, we have an increase of the number of
asperities and a decrease of the mean volume at the same time. If we
denote by $\overline{f}$ the mean of $f$ and by $\overline{v}$ the
mean of the free volume, one must have $\overline{f}\overline{v}=c$,
where $c$ is a constant. Then $\log
\overline{f}=-\log(\overline{v})+\log(c)$, which provides the slope
$-1$ of the line connecting the average values
$(\overline{v},\overline{f})$ at different resolutions in
Fig.\ref{fig12b}. Concerning the amount of the horizontal shift of
the curves by varying $m$, one can assume that
$\overline{v}\propto(L/2^m)^D$. Taking the logarithm of this scaling
law, we obtain $\log(\overline{v})=-Dm\log(2)+\log{c'}$, where $c'$
is a constant. Hence, for $D=2.3$, we expect
$log(\overline{v})=-0.7m+\log(c')$. A translation of $1$ $m$ unity
gives a translation of $0.7$ units in the horizontal coordinate,
which is close to the actual shift of $\overline{v}$ from numerical
data, see for instance $\log(8)-\log(2)=0.6$ by passing from $m=6$
to $m=7$.

\section{Conclusion}

In the present study, the dependency of the free volume between
fractal rough surfaces in contact as a function of the real contact
area and of the contact pressure has been studied by using the
boundary element method. Two main aspects related to the surface
morphological properties have been investigated, namely the effect
of the surface fractal dimension $D$ and the effect of the surface
resolution, which is related to the surface generation parameter $m$
of the random midpoint displacement algorithm.

Examining the relation between the free volume and the real contact
area, we found a nonlinear decay. The free volume is diminished in
case of small surface fractal dimensions (small $D$), or in case of
very refined surfaces (high $m$). Concerning the relation between
the free volume and the contact pressure, a nonlinear decay has been
observed. While an increase in the surface fractal dimension
provides a significant increase in $V^*$ for a given value of $p^*$,
a surface refinement leads to a convergence towards a single curve.
This is a notable behavior which does not take place for other
contact relations, like for $A^*$ vs. $p^*$.

A detailed morphological analysis of the contour plot of the free
volume has also been conducted. It has been found that the free
volumes have a complex spatial distribution over the nominal
cross-section area. The contour domains $\mathbb{D}$ corresponding
to different free volume thresholds $v_{th}$ present a local
dimension $\mathcal{D}$ dependent on the scale of observation,
bounded from below by the fractal dimension of the real contact area
and from above by 2, i.e., by the Euclidean dimension of a flat
surface.

Finally, the fractal properties of the free volume domain have been
exploited in order to derive a formula for the computation of the
free volume. Based solely on the probability distribution function
of the free volumes, it allows a straightforward interpretation of
the surface fractal dimension and surface resolution dependencies
observed in the numerical BEM results.

Further theoretical and numerical research in this field is deemed
to be important, especially as far as the rigorous study of the
fluid-structure interaction is concerned, for instance in the case
of a fluid squeezed among the contacting surfaces. This topic can be
very important in mechanical engineering for a detailed modelling of
lubrication by taking into account roughness, an issue important for
wear. In fact, the development of an appropriate elastomer roughness
in radial lip seals has been found to be dependent on the surface
roughness of the shaft \cite{wear1}. Optimized roughness properties
and texturing can increase the performance of oil seals,
significantly reducing wear \cite{wear2}, and controlling the
direction of leakage flow through the sealing interface. In general,
the control of the seal film thickness, which is directly related to
the free volume between the rough surfaces in contact, can be
beneficial in reducing the forces acting on the mechanical system
and increasing the sealing capabilities \cite{wear3}. In case of
random roughness, an accurate computation of the free volume between
surfaces in contact and of the statistical parameters related to the
valley distributions can be efficiently achieved based on the
computational method herein proposed, fully accounting for the
elastic deformation of the interface. Hence, the proposed method can
be applied to textured surfaces as well, since there are no
restrictions on the form of the height field given in input.

Other technological applications relevant for the present study
regard interfaces in composites for energy applications, like
photovoltaic modules. In those cases, the nonuniform thickness of
polymer seals induced by roughness can be dangerous for the
long-term reliability of these systems exposed to environmental
conditions. Humidity percolation and the evolution of chemical
degradation of the electric contact between Silicon and the
deposited silver conductors is in fact strongly correlated to
imperfect sealing and voids \cite{iseghem}. The characterization of
how leakage mechanisms take place in those systems is of paramount
importance and tailoring of surface roughness may lead to a new
generation of devices with improved lifetime.

\vspace{1em} \addcontentsline{toc}{section}{Acknowledgements}
\noindent\textbf{Acknowledgements} \vspace{1em}

MP would like to thank the Université Paris EST for supporting his
visiting full professorship in the Laboratoire de Modélisation et
Simulation Multi Echelle during May 2014. The research leading to
these results has received funding from the European Research
Council under the European Union's Seventh Framework Programme
(FP/2007-2013) / ERC Grant Agreement n. 306622 (ERC Starting Grant
``Multi-field and multi-scale Computational Approach to Design and
Durability of PhotoVoltaic Modules" - CA2PVM). The authors would
like to thank the anonymous Reviewer for having brought to their
attention the scaling law underlying Fig.12(b).

\bibliographystyle{unsrt}

\end{document}